# Physics-informed neutral network for friction-involved nonsmooth dynamics problems

Zilin Li[a,b,e,1*], Jinshuai Bai[b,1], Huajing Ouyang[d*], Saulo Martelli[b], Jun Zhao[a], Ming Tang[f], Hongtao Wei[a,e], Pan Liu[a,e], Wei-Ron Han[a,c,e*], Yuantong Gu[b*]

[a] *School of Mechanics and Safety Engineering, Zhengzhou University, Zhengzhou 450001, China.*

[b] *School of Mechanical, Medical and Process Engineering, Queensland University of Technology, Brisbane, Australia.*

[c] *Institute of Intelligent Sensing, Zhengzhou University, Zhengzhou 450001, China.*

[d] *School of Engineering, University of Liverpool, Liverpool L19 3GH, U.K.*

[e] *Engineering Technology Research Center of Henan Province for MEMS Manufacturing and Application, Zhengzhou University, Zhengzhou 450001, China.*

[f] *Centre for Genomics and Personalised Health at the Translational Research Institute, School of Biomedical Sciences, Queensland University of Technology, 37 Kent Street, Woolloongabba, QLD 4102, Australia*

## Abstract

Friction-induced vibration (FIV) is widely involved in the engineering. Analysing the dynamic behaviour of the systems containing multiple contact points at the frictional interface is an important topic. However, accurately simulating the repeated nonsmooth/discontinuous dynamic behaviour is challenging. This study presents a new numerical approach for solving nonsmooth friction-induced vibration or friction-involved vibration problems by using the physics-informed neural network (PINN). The neural network training is guided by the theoretical formulations from non-smooth dynamics and multi-body dynamics. Four kinds of high-accuracy PINN-based methods are proposed: (1) single PINN; (2) dual PINN; (3) advanced single PINN; (4) advanced dual PINN. The single PINN uses the nonsmooth dynamic formulations for multibody systems as physical constraints, which is able to substitute the time iteration method in the conventional dynamic analysis. For the dual PINN strategy, in conjunction with the single PINN, a new PINN is proposed as an alternative to traditional numerical methods for solving the linear complementary problems (LCP). The advanced single/dual PINN approaches improve the corresponding single/dual PINN frameworks by incorporating an interpolation technique during the training process. The success of the proposed PINN frameworks in predicting the nonlinear vibration of two typical nonsmooth dynamics problems are verified: one is a 1-dimensional system with direct rigid contact considering stick-slip oscillation, and the other is a 2-dimensional system with spring contact considering separation-reattachment and stick-slip oscillation. The newly proposed PINN frameworks not only illustrate high accuracy in predicting nonsmooth dynamic behaviour, but also eliminate the need for extremely small time steps typically associated with conventional time-stepping methods. Furthermore, PINN methods outperform conventional methods across various friction models.

Key words：Physics-informed neural network, Nonsmooth dynamics, Stick-slip, Contact loss, Friction-induce vibration;

# 1  Introduction

Friction is ubiquitous in daily life as well as in engineering. For centuries, humans have exploited friction and attempted to control friction. When friction is involved in a dynamic environment, the dissipating energy property of friction is generally used to reduce motion or vibration. However, in certain conditions, the stability of a system counterintuitively degrades due to friction [1, 2]. Friction-induced vibration (FIV) [3-5] is one kind of self-excited vibration that associates with a variety of engineering issues, including squealing brakes[3], squeaky joints [6], chattering of cutting tools [7], inaccurate positioning of the robotics [8], etc. The dynamics of the systems with friction are protean and complex. There is a lack of comprehensive understanding of the dynamic characteristics when friction is involved. The manifestations of nonsmoothness that could be directly attributed to friction are: (1) the direction-changing feature of friction force with respect to the relative velocity; (2) stick-slip vibration, which is known as one of the main mechanisms of FIV [9]; (3) separation-reattachment events [6] resulting from unstable FIV. As the contact states change during vibration in a dynamic environment, even systems that are linear within each motion regime (e.g., stick or slip) become nonlinear and time-varying.

Research works on FIV choose to focus on two kinds of systems: (1) low-degrees-of-freedom theoretical models and (2) complicated structures. As for the former kind, models are usually nonlinear and often even nonsmooth. Popp [10] pointed out that even in a one-degree-of-freedom slider-belt model, dry friction-induced stick-slip vibration could be periodic, quasi-periodic and even chaotic. Leine [11] proposed a switching method that is conducive for stick-slip analysis of the slider-belt model with a single contact point. With the development of unstable vibration, loss of contact may occur. Li [12] used a slider-belt system with an inclined spring to reveal that separation and reattachment events could happen when unstable vibration was aroused by mode-coupling instability. In their follow-up work [13], a slider moving on an elastic disc also exhibited stick-slip vibration and the separation and reattachment phenomenon. Generally speaking, the conventional approach for determining FIV of a theoretical model that takes into account nonsmooth features such as stick-slip or separation-reattachment involves tracking the motion states (stick or slip, and contact or separation), which is tedious and time-consuming. For example, Pascal [14] studied the stick-slip motion of a two-degrees-of-freedom lumped-mass-belt model when two sliders were involved, and the stick-slip transition was monitored and captured to ensure the equations of motion with respect to the right motion states were used.

For detailed finite element models of complicate structures, such as automotive brakes, they are often linearised [15] for stability analysis. Such a finite element model can have a large number of contact points. Chen et al. [16] found that in a finite element model with multiple contact points, separation took place between various contact nodes in different regions of the contact interface. However, stick-slip oscillation could not be included in their simulation. It can be found from the published literature that the modelling and solution strategies [11] that are suitable for problems with very few contact points are no longer valid for large contact points. Dealing with the contact and friction when the number of contact points is large has been a difficult problem, which promotes the development of the modelling or solution methods [17-20]. Moreau [52] proposed the differential measurement theory and scanning process theory for nonsmooth dynamic analysis, which is one of the theoretical foundations of the multibody dynamics. The unilateral contact problem of multibody systems can be transformed into linear complementary problem (LCP). A number of methods based on the LCP formulations for dynamic simulations of systems with rigid or elastic unilateral contact have been developed [21-23]. Two general approaches in multibody dynamics are the event-driving method and the time-stepping method [18-20]. For the friction-induced vibration problem with multiple contact points that displays frequent separation-reattachment and stick-slip events, the states of each contact point could influence the entire dynamics of the system and subsequently influence the states of other contact points. The complex nonsmooth dynamics are hard to capture unless a tiny time step is used.

In 2019, Raissi [24] proposed the physics-informed neutral network (PINN) for solving the forward and inverse problems. The physical information, such as partial differential equations, initial conditions and boundary conditions, are used to regularize the deep learning neural network. Karniadakis et al. [25] and Bai et al. [26] reviewed the development of PINN in various fields. Combining the advantages of neural networks and physics laws, PINN is effective for dealing with problems with insufficient data, and thus has the prospect of a wide range of applications. Samaniego [27] presented an improved energetic PINN framework, which was applied to the simulation of hyperelasticity, crack propagation and piezoelectric beams and plates. Gu's team [28, 29] developed a series of computing and modelling methods based on PINN for hydrodynamics and topology optimisation. Pfrommer et al. [30] solved the contact-induced dynamic discontinuous problem of a rigid body by a PINN. A comprehensive loss function involving governing equation, unilateral constraint law and maximum energy dissipation law was defined in their work. The new method could represent realistic impact and stick motion even when spare training data is used.

Accurately predicting discontinuous or nonsmooth dynamic phenomena is a critical challenge in various applications. The development of numerical methods capable of achieving real-time dynamic behaviour in complex systems with friction holds great significance. The primary goal of this research is to introduce a novel methodology that combines the strengths of neural networks and nonsmooth dynamic theories. While using the physics-informed neutral networks to replace the conventional methods, several challenges under the new frameworks need to overcome. This paper is organised as follows: (1) Section 2 first presents the mix-level time-stepping modelling method and then proposes the frameworks of PINN for solving LCP, and the single and dual PINN for the dynamic simulation of the complex systems with friction and contact loss is introduced. (2) Section 3 illustrates the application of proposed single PINN and dual PINN frameworks in the FIV problem with stick-slip. (4) In Section 4, the advanced single/dual PINN frameworks are proposed, which show good accuracy in solving the FIV problem with separation-reattachment and stick-slip.

## 2 Theoretical Approach and Numerical Implementation

### 2.1 Linear complementary problem in unilateral contact

Mathematically, unilateral contact can be described by the linear complementary conditions. Specifically, Fig.1(a) shows the linear complementary relations for single-point rigid contact in the normal direction. The distance between the object of interest and the other contact surface is denoted by $g_{Ni}$, and the contact force by $\lambda_{Ni}$.

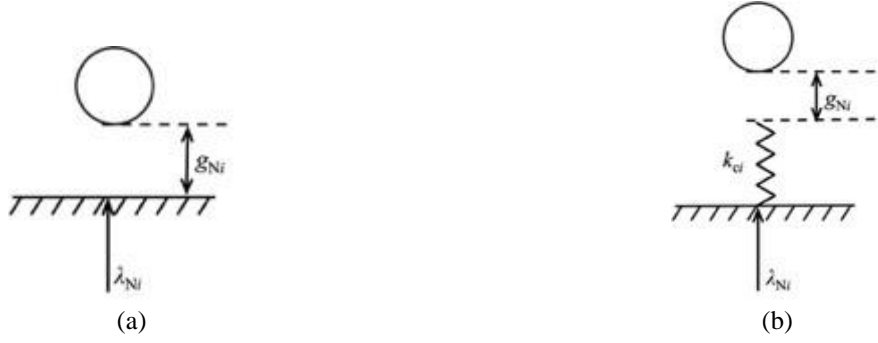

(a)                          (b)

Figure 1. The rigid contact and spring contact.

These two quantities satisfy the relationships that if $g_{Ni} \geq 0$, $\lambda_{Ni}=0$; otherwise if $g_N=0$, $\lambda_{Ni} \geq 0$. For systems with $m$ contact points, $g_{Ni}$ ($i=1, 2, …, m$) is the $i$th element of vector $\mathbf{g}_N$ and $\lambda_{Ni}$ ($i=1, 2,…, m$) is the $i$th element of vector $\boldsymbol{\lambda}_N$. The complementary relationship is given in the vector form:

$$0 \leq \mathbf{g}_N \perp \boldsymbol{\lambda}_N \geq 0 \tag{1}$$

Additionally, for spring contact shown in Fig.1(b), $\Omega_{Ni}=k_c g_{Ni}+\lambda_{Ni}$ is introduced. The following revised relationship now applies: if $\lambda_{Ni}=0$, $\Omega_{Ni} \geq 0$, otherwise if $\lambda_{Ni} \geq 0$, $\Omega_{Ni}=0$. For multiple spring contact, $\Omega_{Ni}$ ($i=1, 2, …, m$) is the $i$th element of vector $\boldsymbol{\Omega}_N$. The complementary relationship is written as:

$$0 \leq \boldsymbol{\Omega}_N \perp \boldsymbol{\lambda}_N \geq 0 \tag{2}$$

When friction is considered in the tangential direction, a general friction law can be described by a function of friction force, namely $\lambda_{Ti}$, with respect to the relative velocity $\gamma_{Ti}$, which is shown in Figure 1 (b). It is generally assumed that sliding friction $\lambda_{Ti}$ follows the Coulomb's law of friction, which is expressed as $\lambda_{Ti}=\mu_i \lambda_{Ni}$. Like the notation rules stated in the normal contact, $\lambda_{Ti}$ and $\gamma_{Ti}$ ($i=1, 2,…, m$) are the elements of vector $\boldsymbol{\lambda}_T$ and $\boldsymbol{\gamma}_T$ for describing the friction and relative velocity of the multi-point contact problem. Through mathematical transformation, the friction law can be decomposed into two complementary relations:

$$0 \leq \boldsymbol{\lambda}_R^T \perp \boldsymbol{\gamma}_R \geq 0, \quad 0 \leq \boldsymbol{\lambda}_L^T \perp \boldsymbol{\gamma}_L \geq 0 \tag{3}$$

where $\boldsymbol{\gamma}_R = \frac{1}{2}(|\boldsymbol{\gamma}_T|+\boldsymbol{\gamma}_T)$, $\boldsymbol{\gamma}_L = \frac{1}{2}(|\boldsymbol{\gamma}_T|-\boldsymbol{\gamma}_T)$, $\boldsymbol{\lambda}_R = \boldsymbol{\mu}\boldsymbol{\lambda}_N + \boldsymbol{\lambda}_T$ and $\boldsymbol{\lambda}_L = \boldsymbol{\mu}\boldsymbol{\lambda}_N - \boldsymbol{\lambda}_T$,

with the relationship

$$\boldsymbol{\gamma}_T = -\boldsymbol{\gamma}_R + \boldsymbol{\gamma}_L \quad \text{and} \quad \boldsymbol{\lambda}_R = 2\boldsymbol{\mu}\boldsymbol{\lambda}_N - \boldsymbol{\lambda}_L \tag{4}$$

or between the velocity and impulse of the friction force:

$$0 \leq \mathbf{\Lambda}_R^T \perp \boldsymbol{\gamma}_R \geq 0, \quad 0 \leq \mathbf{\Lambda}_L^T \perp \boldsymbol{\gamma}_L \geq 0 \tag{5}$$

where $\mathbf{\Lambda}_R = \int \boldsymbol{\lambda}_R dt$ and $\mathbf{\Lambda}_L = \int \boldsymbol{\lambda}_L dt$

## 2.2 Equation of motion of the multibody system with unilateral contact

For the dynamic modelling of the multibody systems with unilateral contact, the equation of motion in matrix form is generally written as Eq.(6):

$$\mathbf{M}\ddot{\mathbf{q}} - \mathbf{h} - \mathbf{W}_N \boldsymbol{\lambda}_N - \mathbf{W}_T \boldsymbol{\lambda}_T = \mathbf{0} \tag{6}$$

in which $\mathbf{M}$ is the mass matrix, $\mathbf{h}$ is the force vector for the system regardless of the unilateral conditions, $\boldsymbol{\lambda}_N$ and $\boldsymbol{\lambda}_T$ are the normal and tangential force vectors, respectively. $\mathbf{W}_N$ is the displacement vector defined in the global coordinates. The constraint functions in the normal direction and tangential direction are expressed as $\mathbf{g}_N = \mathbf{W}_N^T \mathbf{q}$, and $\mathbf{g}_T = \mathbf{W}_T^T \mathbf{q}$. $\mathbf{W}_N$ and $\mathbf{W}_T$ can be obtained by $\mathbf{W}_N = \left(\frac{\partial \mathbf{g}_N}{\partial \mathbf{q}}\right)^T$ and $\mathbf{W}_T = \left(\frac{\partial \boldsymbol{\gamma}_T}{\partial \mathbf{u}}\right)^T$.

Then, take the first derivative of $\mathbf{g}_N$ and $\boldsymbol{\gamma}_T$ with respect to time, the normal velocity and tangential acceleration can be expressed as:

$$\boldsymbol{\gamma}_N = \mathbf{W}_N^T \dot{\mathbf{q}} + \hat{\boldsymbol{\omega}}_N, \quad \dot{\boldsymbol{\gamma}}_T = \mathbf{W}_T^T \ddot{\mathbf{q}} + \frac{\partial \mathbf{W}_T^T}{\partial t}\dot{\mathbf{q}} + \dot{\hat{\boldsymbol{\omega}}}_T \tag{7}$$

in which $\hat{\boldsymbol{\omega}}_N = \frac{\partial \mathbf{W}_N^T}{\partial t}$ and $\dot{\hat{\boldsymbol{\omega}}}_T = \frac{\partial^2 \mathbf{W}_T^T}{\partial t^2}$.

By discretising the motion states over time interval $\Delta t = t_E - t_A$, one obtained the time-domain discretised equation of motion and the state vector:

$$\begin{cases} \mathbf{M}^{(A)} \Delta \mathbf{u} - \mathbf{h}^{(A)} \Delta t - \mathbf{W}_N^{(A)} \mathbf{\Lambda}_N - \mathbf{W}_T^{(A)} \mathbf{\Lambda}_T = \mathbf{0} \\ \Delta \mathbf{q} = (\mathbf{u}^{(A)} + \Delta \mathbf{u}) \Delta t \end{cases} \tag{8}$$

in which $\mathbf{\Lambda}_N = \boldsymbol{\lambda}_N \Delta t$, $\mathbf{\Lambda}_T = \boldsymbol{\lambda}_T \Delta t$ and superscript (A) represents the corresponding physical quantity at instant $t_A$.

The increment of constraint function $\Delta \mathbf{g}_N$ and $\Delta \boldsymbol{\gamma}_T$ over $\Delta t$ can be obtained according to Eq.(7):

$$\Delta \mathbf{g}_N = \mathbf{W}_N^{(A)T} \Delta \mathbf{q} + \hat{\boldsymbol{\omega}}_N \Delta t, \quad \Delta \boldsymbol{\gamma}_T = \mathbf{W}_T^T \Delta \mathbf{u} + \frac{\partial \mathbf{W}_T^T}{\partial \mathbf{q}} \Delta \mathbf{q} + \dot{\hat{\boldsymbol{\omega}}}_T \Delta t \tag{9}$$

In corporation with Eq.(8), motion states and the constraint function at the next time instance $t_E$ are derived:

$$\mathbf{g}_N^{(E)} = \mathbf{G}_{NN}\mathbf{\Lambda}_N^{(E)}\Delta t + \mathbf{G}_{NT}\mathbf{\Lambda}_T^{(E)}\Delta t + \mathbf{g}_N^{(A)} + \mathbf{W}_N^{(A)T}\mathbf{u}^{(A)}\Delta t + \mathbf{G}_N\mathbf{h}^{(A)}\Delta t + \hat{\boldsymbol{\omega}}_N\Delta t \tag{10}$$

$$\boldsymbol{\gamma}_T^{(E)} = \mathbf{G}_{TN}\mathbf{\Lambda}_N^{(E)} + \mathbf{G}_{TT}\mathbf{\Lambda}_T^{(E)} + \dot{\mathbf{g}}_T^{(A)} + \mathbf{G}_T\mathbf{h}^{(A)}\Delta t + \frac{\partial \mathbf{W}_T^{(A)T}}{\partial \mathbf{q}}\mathbf{u}^{(A)}\Delta t + \dot{\hat{\boldsymbol{\omega}}}_T\Delta t \tag{11}$$

in which $\mathbf{G}_N = \mathbf{W}_N^{T(A)}\mathbf{M}^{-1(A)}$, $\mathbf{G}_{NN} = \mathbf{W}_N^{T(A)}\mathbf{M}^{-1(A)}\mathbf{W}_N^{T(A)}$, $\mathbf{G}_{NT} = \mathbf{W}_N^{T(A)}\mathbf{M}^{-1(A)}\mathbf{W}_T^{T(A)}$,

$\mathbf{G}_T = \mathbf{W}_T^{T(A)}\mathbf{M}^{-1(A)} + \frac{\partial \mathbf{W}_T^T}{\partial \mathbf{q}}\mathbf{M}^{-1(A)}\Delta t$, $\mathbf{G}_{TN} = \mathbf{G}_T\mathbf{W}_N^{T(A)}$, $\mathbf{G}_{TT} = \mathbf{G}_T\mathbf{W}_T^{T(A)}$.

There are six unknown vectors in the four equations in Eqs. (8), (10) and (11). To solve the problem, another two complementary equations need to be introduced. In the following, the theoretical formulations for two contact conditions are presented.

2.2.1 Time-stepping formulations for the unilateral rigid contact at displacement-velocity level

As stated in Section 2.1, unilateral rigid contact can be enforced using a complementary relation between $g_N$ and $\lambda_N$. The friction is independent of the state solutions in the normal direction, which follow the complementary relation given in Eq.(1). According to Eq.(4), $\boldsymbol{\gamma}_T^{(E)}$ and $\mathbf{\Lambda}_T^{(E)}$ in Eq.(11) can be replaced by $\boldsymbol{\gamma}_L^{(E)}$ and $\mathbf{\Lambda}_R^{(E)}$. And then, by introducing $\mathbf{\Lambda}_R = 2\boldsymbol{\mu}\mathbf{\Lambda}_N - \mathbf{\Lambda}_L$ according to Eq.(4), the LCP equations in the matrix form can be obtained:

$$\underbrace{\begin{Bmatrix} \mathbf{g}_N^{(E)} \\ \boldsymbol{\gamma}_L^{(E)}\Delta t \\ \mathbf{\Lambda}_R^{(E)}\Delta t \end{Bmatrix}}_{\mathbf{y}} = \underbrace{\begin{bmatrix} \mathbf{G}_{NN}+\mathbf{G}_{NT}\boldsymbol{\mu} & -\mathbf{G}_{NT} & \mathbf{0} \\ -(\mathbf{G}_{TN}+\mathbf{G}_{TT}\boldsymbol{\mu}) & \mathbf{G}_{TT} & \mathbf{I} \\ 2\boldsymbol{\mu} & -\mathbf{I} & \mathbf{0} \end{bmatrix}}_{\mathbf{A}} \underbrace{\begin{Bmatrix} \mathbf{\Lambda}_N^{(E)}\Delta t \\ \mathbf{\Lambda}_L^{(E)}\Delta t \\ \boldsymbol{\gamma}_R^{(E)}\Delta t \end{Bmatrix}}_{\mathbf{x}} + \underbrace{\begin{Bmatrix} \mathbf{g}_N^{(A)}+\mathbf{W}_N^{(A)T}\Delta t\mathbf{u}^{(A)}+\mathbf{G}_N\mathbf{h}^{(A)}\Delta t+\hat{\boldsymbol{\omega}}_N\Delta t \\ -\boldsymbol{\gamma}_T^{(A)}\Delta t - \left(\mathbf{G}_T\mathbf{h}^{(A)} + \frac{\partial \mathbf{W}_T^T}{\partial \mathbf{q}}\mathbf{u}^{(A)} + \dot{\hat{\boldsymbol{\omega}}}_T\right)\Delta t^2 \\ \mathbf{0} \end{Bmatrix}}_{\mathbf{h}} \tag{12}$$

with

$$0 \leq \mathbf{g}_N^{(E)} \perp \mathbf{\Lambda}_N^{(E)} \geq 0,\ 0 \leq \boldsymbol{\gamma}_L^{(E)} \perp \mathbf{\Lambda}_L^{(E)} \geq 0 \text{ and } 0 \leq \boldsymbol{\gamma}_R^{(E)} \perp \mathbf{\Lambda}_R^{(E)} \geq 0$$

2.2.2 Time-stepping formulations for the unilateral spring contact at the displacement-velocity level

In contrast with direct contact, the complementary condition of the spring contact is between $\boldsymbol{\Omega}_N$ and $\lambda_N$. Meanwhile, the friction force depends on the state solutions in the normal direction when contact is maintained. By substituting Eq. (9) into $\boldsymbol{\Omega}_N$, one gets:

$$\boldsymbol{\Omega}_{NE} = \left(\mathbf{I}/\Delta t^2 + \mathbf{k}_C \mathbf{G}_{NN} + \mathbf{k}_C \mathbf{G}_{NT}\boldsymbol{\mu}\right)\boldsymbol{\Lambda}_N \Delta t - \mathbf{k}_C \mathbf{G}_{NT}\boldsymbol{\Lambda}_L \Delta t + \mathbf{k}_C \left(\mathbf{g}_{NA} + \mathbf{W}_{NA}^T \Delta t \mathbf{u}_A + \mathbf{G}_N \mathbf{h}_A \Delta t + \hat{\boldsymbol{\omega}}_N \Delta t\right) \quad (13)$$

Collecting Eqs. (10), (13) and $\boldsymbol{\Lambda}_R = 2\boldsymbol{\mu}\boldsymbol{\Lambda}_N - \boldsymbol{\Lambda}_L$ in matrix form gives the LCP formulations of the whole system:

$$\underbrace{\begin{Bmatrix} \boldsymbol{\Omega}_N^{(E)} \\ \boldsymbol{\gamma}_L^{(E)} \Delta t \\ \boldsymbol{\Lambda}_R^{(E)} \Delta t \end{Bmatrix}}_{\mathbf{y}} = \underbrace{\begin{bmatrix} \mathbf{I}/\Delta t^2 + \mathbf{k}_C \mathbf{G}_{NN} + \mathbf{k}_C \mathbf{G}_{NT}\boldsymbol{\mu} & -\mathbf{k}_C \mathbf{G}_{NT} & \mathbf{0} \\ -(\mathbf{G}_{TN} + \mathbf{G}_{TT}\boldsymbol{\mu}) & \mathbf{G}_{TT} & \mathbf{I} \\ 2\boldsymbol{\mu} & -\mathbf{I} & \mathbf{0} \end{bmatrix}}_{\mathbf{A}} \underbrace{\begin{Bmatrix} \boldsymbol{\Lambda}_N^{(E)} \Delta t \\ \boldsymbol{\Lambda}_L^{(E)} \Delta t \\ \boldsymbol{\gamma}_R^{(E)} \Delta t \end{Bmatrix}}_{\mathbf{x}} + \underbrace{\begin{Bmatrix} \mathbf{k}_C \left(\mathbf{g}_N^{(A)} + \mathbf{W}_N^{(A)T}\mathbf{u}^{(A)}\Delta t + \mathbf{G}_N \mathbf{h}^{(A)}\Delta t + \hat{\boldsymbol{\omega}}_N \Delta t\right) \\ -\boldsymbol{\gamma}_T^{(A)}\Delta t - \left(\mathbf{G}_T \mathbf{h}^{(A)} + \frac{\partial \mathbf{W}_T^T}{\partial \mathbf{q}}\mathbf{u}^{(A)} + \dot{\hat{\boldsymbol{\omega}}}_T\right)\Delta t^2 \\ \mathbf{0} \end{Bmatrix}}_{\mathbf{h}} \quad (14)$$

with

$$0 \leq \boldsymbol{\Omega}_N^{(E)} \perp \boldsymbol{\Lambda}_N^{(E)} \geq 0,\ 0 \leq \boldsymbol{\gamma}_L^{(E)} \perp \boldsymbol{\Lambda}_L^{(E)} \geq 0 \text{ and } 0 \leq \boldsymbol{\gamma}_R^{(E)} \perp \boldsymbol{\Lambda}_R^{(E)} \geq 0$$

Eventually, Eqs. (8) and (14) together form the equations for the nonsmooth dynamic analysis of the complex frictional system with multi-contact points. The mix-level time-stepping method is called as the conventional method in the following part.

## 2.3 Physics-informed neutral network for nonsmooth dynamics

Compared with the event-detection strategy, the time-stepping strategy is more suitable for problems with a considerable number of contact points as it avoids the repeated detection of the state transition. However, as precision cannot be compromised, a tiny time step is required for solving complex nonlinear vibration. Thus, the expense of computation becomes tremendous.

In order to achieve high-accuracy calculation of nonsmooth dynamics, in this work, a new approach is proposed by utilising the physics-informed neural network (PINN). A PINN consists of the feedforward neural network (FNN) and a physics-informed loss function [26], as shown in Figure 2. In a PINN, the FNN is used to capture the relationship between the input data and the output data, while the physics-informed loss function embeds the physics laws and quantifies the performance of the FNN [31].

As observed from Figure 2, an FNN comprises three parts, namely the input layer, the hidden layers and the output layer. When using an FNN, input data are fed into the input layer and forwardly transported to the next layer. Eventually, the final predictions of an FNN are output from the output layer. An L-layer FNN can be mathematically expressed as

$$\begin{aligned} {}^0\mathbf{a} &= {}^0\mathbf{a}, \\ {}^l\mathbf{a} &= \sigma\left({}^{l-1}\mathbf{w} \cdot {}^{l-1}\mathbf{a} + {}^{l-1}\mathbf{b}\right), \\ {}^L\mathbf{a} &= {}^{L-1}\mathbf{w} \cdot {}^{L-1}\mathbf{a}, \end{aligned} \tag{15}$$

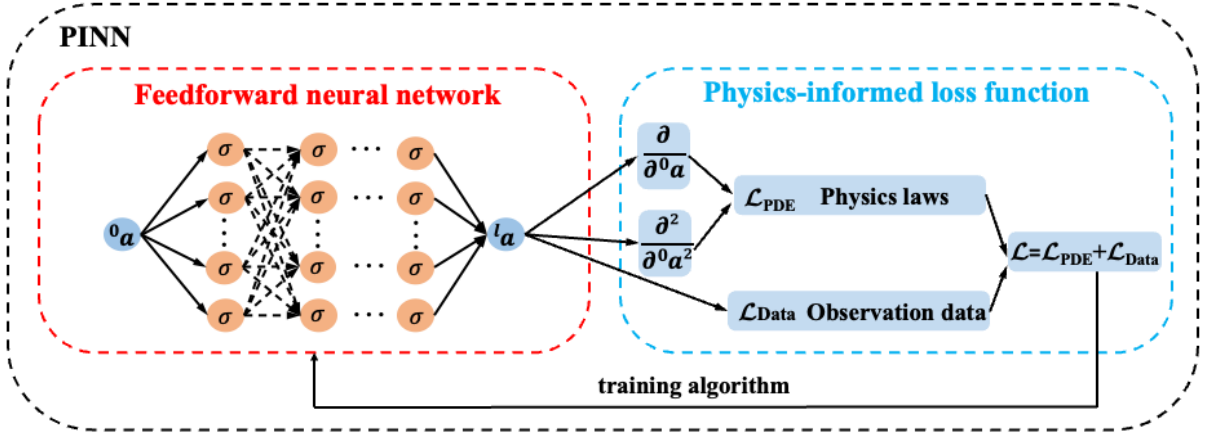

Figure 2. An example of a physics-informed neural network (PINN).

Where $^0\mathbf{a}$ and $^L\mathbf{a}$ denote the input layer (the $0^{th}$ layer) and the output layer (the $L^{th}$ layer), respectively. $l = 1, 2, \ldots, L-1$ denotes the number of hidden layers. $\mathbf{w}$ and $\mathbf{b}$ are the weights and biases in the FNN. $\sigma$ denotes the activation function that adds nonlinearity to the FNN [32]. More details regarding the activation function are given in the following context. On the right-hand side of Figure 2, the outputs of the FNN are used to formulate the physics-informed loss function. It is worth noting that, when solving PDEs (partial differential equations) via FNNs, the partial differential terms in the target PDEs can be analytically obtained through automatic differentiation [33]. Through training algorithms [34], the parameters of the FNN are iteratively modified in order to decrease the value of the physics-informed loss function. Finally, the training process converges when the given criteria are satisfied.

To calculate the nonsmooth dynamic problem, two fundamental solution problems are involved, namely, the solution of the LCP equation and the calculation of the dynamic response. In what follows, different neural network structures for the LCP equation and dynamics response are introduced in detail, respectively.

### 2.3.1 *PINN for* solving LCP equations

This part presents the idea of combining the neural network and the theoretical description of LCP. The general LCP equation is in the form of Eq. (16):

$$\begin{cases} \mathbf{y} = \mathbf{Ax} + \mathbf{b} \\ \mathbf{y}^T \mathbf{x} = \mathbf{0} \\ \mathbf{x} \geq \mathbf{0}, \mathbf{y} \geq \mathbf{0} \end{cases} \tag{16}$$

in which **x** and **y** are vectors.

Instead of the conventional methods (pivoting or the quadratic programming method (QP method)) to get the solutions, we define two functions:

$$\mathbf{f} = \mathbf{y} - \mathbf{Ax} + \mathbf{b} \tag{17}$$

$$\mathbf{r} = \mathbf{y}^T \mathbf{x} \tag{18}$$

The loss function in the following form is defined:

$$\mathcal{L} = \frac{\sum_{i=1}^{N} f_i^2 + \sum_{i=1}^{N} r_i^2}{N} \tag{19}$$

in which $f_i$ and $r_i$ are respectively the element in vector **f** and **r**, and $N$ is the element number of the vector. Physics-informed loss function $\mathcal{L}$ serves as the optimisation object in the training process of the neural network.

The architecture of the PINN for solving the LCP equation (Eq. (17)) is shown in Figure 3.

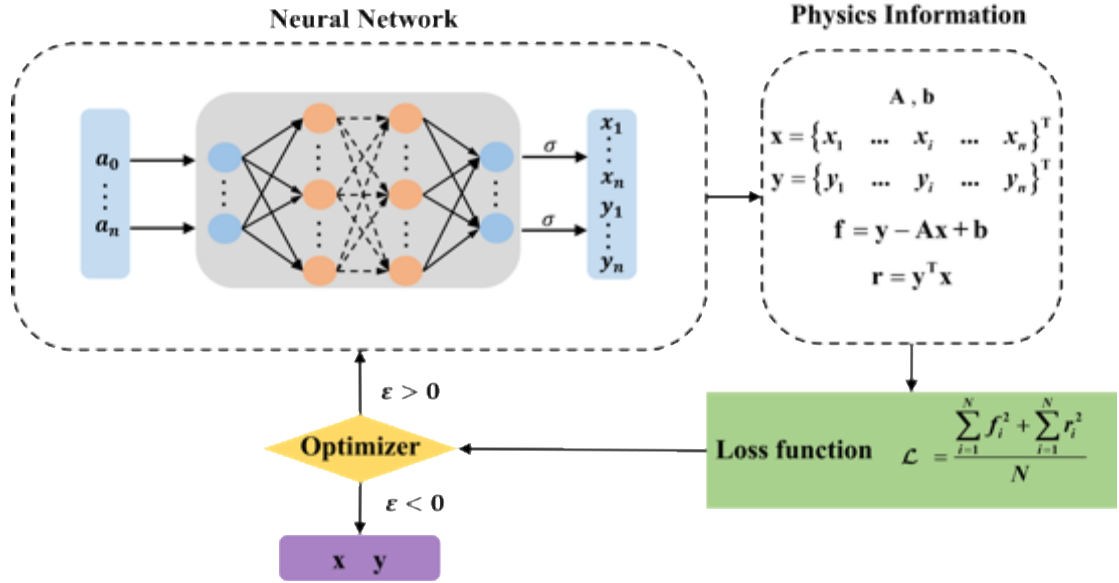

Figure 3. The process of the PINN framework for LCP.

By starting with $a_0$, $a_1$, …, $a_n$ as the input, the neural network is used for predicting the output data $x$ and $y$ data. In this procedure, several variables need to be determined. One critical choice is the activation function between the layers. Several nonlinear activation functions can be used, such as the tanh, mish and ReLU, shown in Figure 4.

As $x$ and $y$ in LCP need to be either positive or zero values, the ReLU function given in Figure 3 ()can naturally provide such an output satisfying the LCP condition. To further improve the output values, a new activation function is designed based on the ReLU function, which is in the following form:

$$g = \max(0, a + c_1) + c_2 \qquad (20)$$

where $c_1$ and $c_2$ are the correction factors. The modified ReLU function could adjust the data near the origin point, shown in Figure 4 (d), which largely save the training time if the machine learning algorithm locates the data at the negative axis.

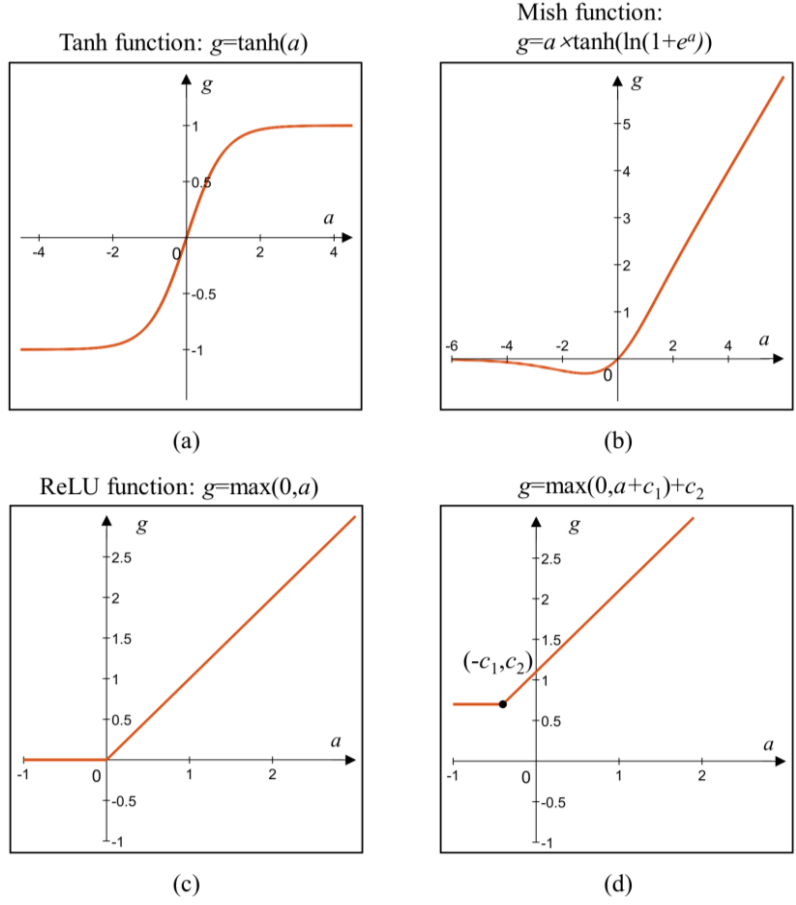

Figure 4. The nonlinear activation functions: (a) Tanh function; (b) Mish function; (c) ReLu function; (d) the modified ReLu function.

After obtaining the outputs $x$ and $y$, the optimisation of the physics-informed loss function is accomplished by L-BFGS algorithm. As shown in the overall procedure of the LCP PINN (Figure 3), the training ends when the error becomes smaller than the tolerance.

Several examples are calculated by the LCP PINN framework as well as two conventional methods (Pivoting and QP method) for comparison, with $A = \begin{bmatrix} 1 & -1 \\ -1 & 0 \end{bmatrix}$ and $b = \begin{bmatrix} -0.009 & 0.02 \end{bmatrix}^T$. For the sake of conciseness of the paper, the solutions are listed in Table 1. This example uses a 3-layer neural network with 5 neutrons at each layer. It is proved that the proposed LCP PINN framework can give accurate solutions.

Table 1 The solutions by different methods

|   | Accurate result | PINN framework | Pivoting method | QP method |
|---|---|---|---|---|
| **x** | $[0.009 \quad 0]^T$ | $[0.009 \quad 0]^T$ | $[0.009 \quad 0]^T$ | $[0.0090004 \quad 4.8204583\text{e-}09]^T$ |
| y | [0  0.011]$^T$ | [0  0.011]$^T$ | [0  0.011]$^T$ | [4.0007060e-07  0.0109996]$^T$ |

2.3.2  *PINN for vibration of multibody systems with unilateral contact*

For the initial condition problem, the second-order differential equation is general in the form $\ddot{q} = f(t, q, \dot{q})$, and the initial conditions are $q(t_0) = q_0$ and $\dot{q}(t_0) = \dot{q}_0$. Implicit Runge-Kutta method is an high-order integral method to solve the initial condition problem. The $R^{\text{th}}$-order implicit Runge-Kutta method can be expressed as a transformation form, given in Eqs. (21) and (22):

$$\dot{q}_k^{(i)} = \dot{q}_k^{(i+c_k)} - \Delta t \sum_{r=1}^{R} a_{k,r} f\left(t^{(i)} + c_r \Delta t, \dot{q}_r^{(i+c_r)}\right), \quad k = 1, 2, \ldots R \tag{21}$$

$$\dot{q}^{(i)} = \dot{q}^{(i+1)} - \Delta t \sum_{k=1}^{R} b_k f\left(t^{(i)} + c_k \Delta t, \dot{q}_k^{(i+c_k)}\right) \tag{22}$$

in which $\Delta t$ is one time increment over discrete time, $a_{k,r}$, $b_k$ and $c_r$ are the coefficients. In Eqs. (21) and (22), the known term at time step *I* are expressed by the unknown terms at time step *i*+1. When a neutral network is introduced, using $q$ as the input data, [ $\dot{q}_k^{(i+c_1)}$, $\dot{q}_k^{(i+c_2)}$,…, $\dot{q}_k^{(i+1)}$ ] can be predicted which is designed as the output data. Subsequently, the neural network is trained until $\dot{q}_r^{(i)}$ and $\dot{q}^{(i)}$ obtained by Eqs. (21) and (22) converge to the corresponding values that has been known.

As to multi-degree-of-freedom systems, the general form of the equation of motion is expressed as:

$$\mathbf{M}\ddot{\mathbf{q}} + \mathbf{C}\dot{\mathbf{q}} + \mathbf{K}\mathbf{q} = \mathbf{f}_e \tag{23}$$

in which **M**, **K**, **C** are the mass, stiffness and damping matrix, **q** is the displacement vector and $\mathbf{f}_e$ is the external force. **v** refers to as $\dot{\mathbf{q}}$. The conventional way to solve this second-order differential equation of dynamical systems is using numerical integral methods, such as the explicit/implicit Runge-Kutta (RK) methods. Here, Figure 5 shows the framework of a PINN for dynamic simulation of multi-degree-of-freedom systems. Although the time integral idea of the current PINN is based on the implicit R-K method but as introduced above, it is actually different from conventional implicit R-K method.

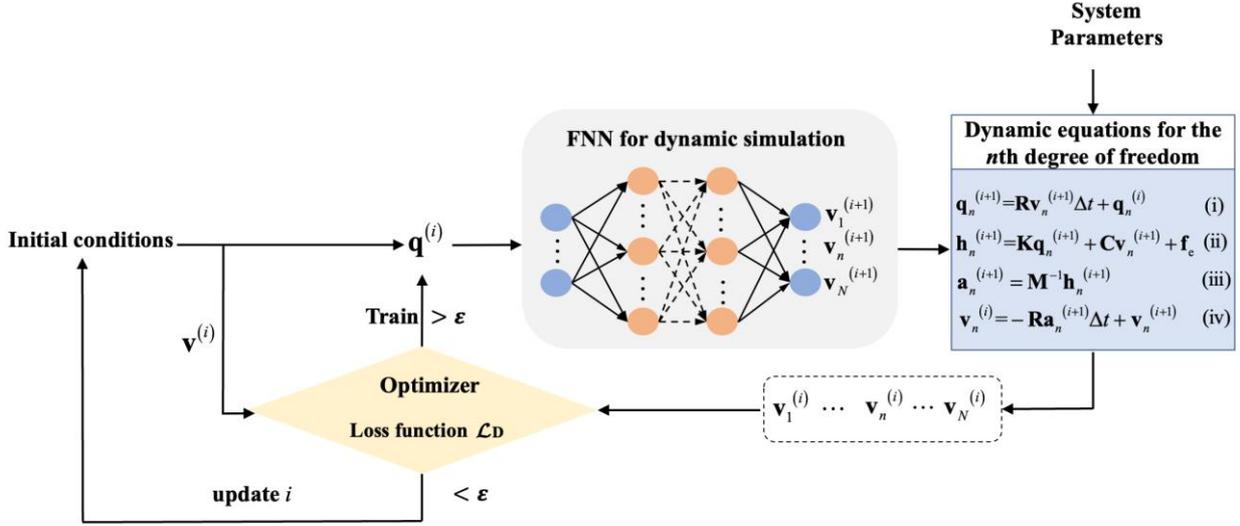

Figure 5. The framework of PINN for dynamic simulation.

In the above framework, the input data are the elements of $\mathbf{q}=[\,q_1,\ldots q_l,\ldots,q_N\,]^T$ ($n=1,2,\ldots,N$) and the network is designed to output $N\times R$ number of data. $N$ is the number of the system degrees of freedom, and $r$ is the number of orders used for the time integral. Then the output data is reshaped to $N$ vectors that each has $R$ elements. Each vector is assigned to the velocity $\mathbf{v}_r^{(i+1)}=[\dot{q}_r^{(i+c_1)},\dot{q}_r^{(i+c_2)},\ldots,\dot{q}_r^{(i+1)}]^T$ corresponding to $r$ number of quantities of each degree-of-freedom. After obtaining the physics quantities $\mathbf{v}_N^{i+1}$, the displacement and the acceleration at time step $i+1$ can be obtained through the dynamic equations (Eqs. (i)-(iii)) shown in Figure 5. $\mathbf{R}$ is the coefficients matrix about $a_{k,r}$ and $b_r$. Thus, according to the idea of Eqs. (21) and (22), which is written as a matrix form Eq. (iv) in Figure 5, the velocity vectors $\mathbf{v}_n^{(i)}=[v_{1,n},\ldots,v_{r,n},\ldots,v_{R+1,N}]$ can be obtained. The displacement and velocity can be learned by optimizing the loss function in the following form:

$$\mathcal{L}_D = \sum_{n=1}^{N}\sum_{r=1}^{R+1}\left(v_{r,n}-v^{(i)}\right) \quad (24)$$

The coefficient tables from the 2th-order to 200th-order implicit Runge-Kutta method can be obtained in https://github.com/maziarraissi/PINNs [24].

### 2.3.3 *Single PINN for the transient dynamic analysis of nonsmooth vibration*

In this part, a new physics-informed neural network that introduces nonsmooth multibody dynamic theory into the neutral network is presented. When unilateral contact is considered, nonsmooth characteristics are involved in the dynamic system. Figure 6 shows the framework of the specific

PINN, which is referred to as the single PINN framework in the following. Compared with the PINN for dynamic simulation given in Figure 5, the equations of motion Eqs. (b) and (c) used in the neural network are:

$$\mathbf{h}_n^{(i+1)} = \mathbf{K}_s \mathbf{q}_n^{(i+1)} + \mathbf{C}_s \mathbf{v}_n^{(i+1)} + \mathbf{f}_e \qquad (25)$$

$$\mathbf{a}_n^{(i+1)} = \mathbf{M}^{-1}\mathbf{h}_n^{(i+1)} + \mathbf{W}_N \lambda_N + \mathbf{W}_T \lambda_T \qquad (26)$$

in which $\mathbf{K}_s$ and $\mathbf{C}_s$ are the stiffness and damping matrix when the unilateral constraints are removed, and the contact force $\lambda_N$ and friction force $\lambda_T$ are time-varying variables, which can be determined by the LCP equations introduced in Section 2.2.1 or 2.2.2.

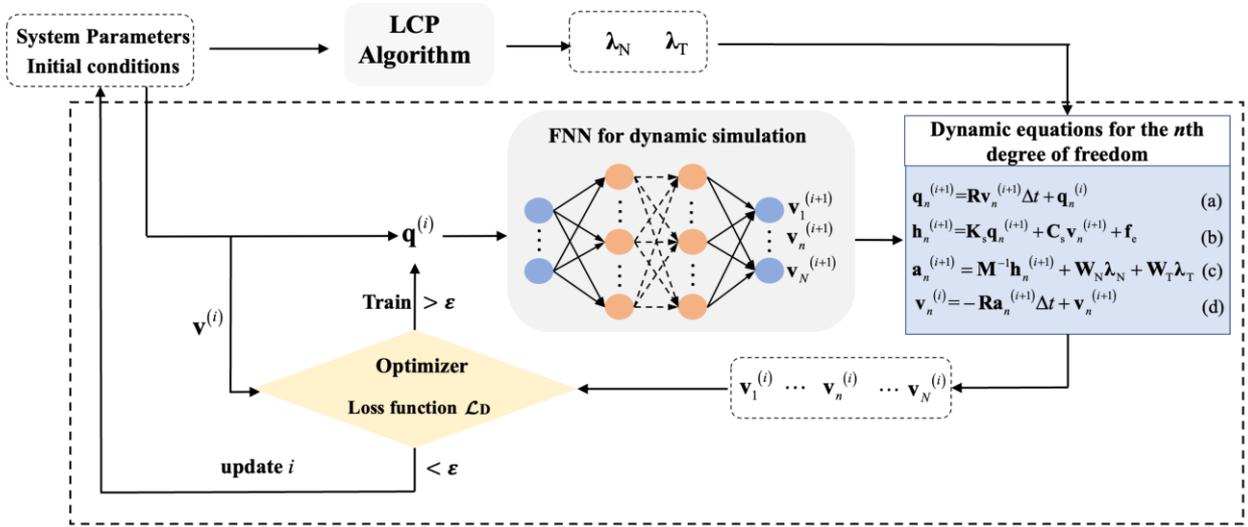

Figure 6. The framework of the single PINN framework.

The calculation process of the single PINN framework is: (1) use the system information, such as the system parameters and the initial conditions to calculate $\lambda_N$ and $\lambda_T$, based on the time-stepping method according to the contact type; (2) transfer $\lambda_N$ and $\lambda_T$ to the dynamic equation part in the PINN for dynamic calculation. The operation mechanism of the PINN for dynamic calculation has been explained in previous section. In terms of the neural network design, the activation function for the dynamic calculation is tanh. The dynamic response can be learned by optimising the loss function $\mathcal{L}_D$ given in Eq.(24).

## 2.3.4 *Dual PINN for the transient dynamic analysis of nonsmooth vibration*

One the basis of the LCP PINN for linear complementary problem proposed in Section 2.3.1 and the PINN for multibody systems with unilateral contact, a new physics-informed neutral network is proposed, which is shown in Figure 7. In this method, two physics-informed networks are involved. Thus, this method is referred to as the dual PINN framework.

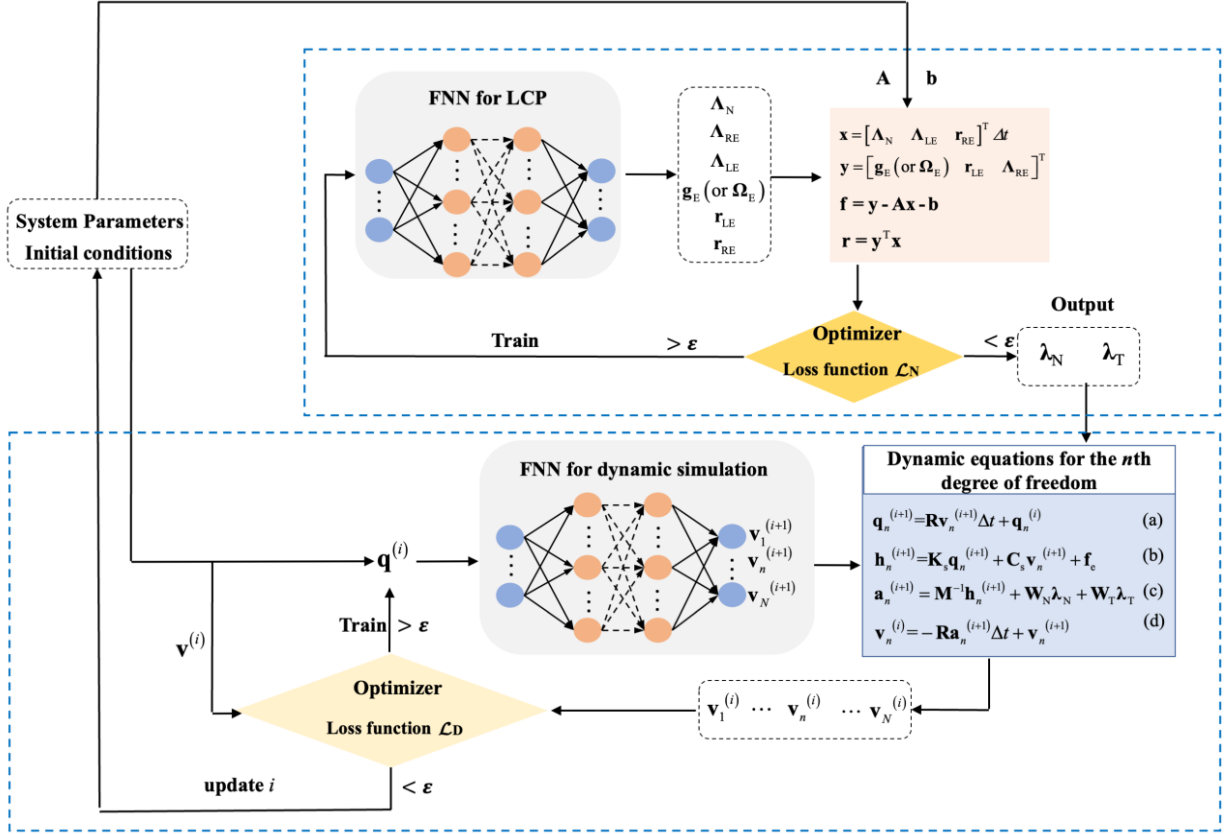

Figure 7. The framework of the Dual PINN.

As the normal contact and tangential contact can be enforced through LCP, the first neutral network is basically the LCP PINN. Compared with the PINN for LCP in Section 2.3.1, the outputs of the FNN for LCP are assigned to physical variables. It should be noticed that for different contact types, the physical variables are different: (1) for direct rigid contact (Section 2.2.1), outputs are assigned to $\Lambda_N$, $\Lambda_{RE}$, $\Lambda_{LE}$, $\mathbf{g}_E$, $\mathbf{r}_{LE}$ and $\mathbf{r}_{RE}$; (2) for rigid contact with springs (Section 2.2.2), outputs are assigned to $\Lambda_N$, $\Lambda_{RE}$, $\Lambda_{LE}$, $\Omega_E$, $\mathbf{r}_{LE}$ and $\mathbf{r}_{RE}$. Then $\mathbf{x} = [\Lambda_N \quad \Lambda_{LE} \quad \mathbf{r}_{RE}]^T \Delta t$, and $\mathbf{y} = [\mathbf{g}_E \quad \mathbf{r}_{LE} \quad \Lambda_{RE}]^T$ (or $\mathbf{y} = [\Omega_E \quad \mathbf{r}_{LE} \quad \Lambda_{RE}]^T$) are defined. For the direct rigid contact, the expression of $\mathbf{A}$ and $\mathbf{b}$ are given in Eq.(12), while for contact with springs, the expression of **A** and **b** are given in Eq. (14). Rather than applying the conventional time-stepping method, the equation of the LCP problem (Eq. (13)) no longer needs to be solved, as all the unknown quantities $\Lambda_N, \Lambda_{RE}, \Lambda_{LE}$, $\mathbf{g}_E$ (or $\Omega_E$), $\mathbf{r}_{LE}$ and $\mathbf{r}_{RE}$ in Eq. (12) or (14) have been predicted by the neural network. Contact force $\lambda_N$ and friction force $\lambda_T$ can be learned by optimising the loss function $L$ (Eq.(19)). The input data of the LCP PINN can be arbitrary values.

When contact force $\lambda_N$ and friction force $\lambda_T$ are outputted, they are transferred to the physics part of the second PINN for the dynamic analysis of the multibody system with unilateral contact. The basic idea of this PINN has been introduced in Section 2.3.4. Eventually, the dynamic response during an arbitrary time duration can be learned through training.

## 3 One-dimensional stick-slip vibration with direct contact

Stick-slip is an essential mechanism of friction-induced vibration and exists in many applications. This section presents the application of PINN frameworks to solve the stick-slip vibration of Figure 8. It is assumed that the contact in the normal direction is maintained during vibration. The vertical position of the slider mass is a constant value of $0$. $\lambda_N$ is known in advance, which is the compression force. The equations of motion for systems with friction contact can be reduced to:

$$\mathbf{M\ddot{x}} + \mathbf{h} + \mathbf{W}_T \lambda_T = \mathbf{0} \tag{27}$$

The LCP equation given in Eq.(12) is degenerated to:

$$\underbrace{\begin{Bmatrix} \gamma_L^{(E)} \\ \Lambda_R^{(E)} \Delta t \end{Bmatrix}}_{\mathbf{y}} = \underbrace{\begin{bmatrix} \mathbf{G}_{TT} & \mathbf{I} \\ -\mathbf{I} & \mathbf{0} \end{bmatrix}}_{\mathbf{A}} \underbrace{\begin{Bmatrix} \Lambda_L^{(E)} \\ \gamma_R^{(E)} \end{Bmatrix}}_{\mathbf{x}} + \underbrace{\begin{Bmatrix} -(\mathbf{G}_{TN} + \mathbf{G}_{TT}\boldsymbol{\mu})\Lambda_N^{(E)} - \gamma_T^{(A)} - \left( \mathbf{G}_T \mathbf{h}^{(A)} + \frac{\partial \mathbf{W}_T^T}{\partial \mathbf{q}} \mathbf{u}^{(A)} + \dot{\boldsymbol{\omega}}_T \right) \Delta t \\ 2\boldsymbol{\mu}\Lambda_N^{(E)} \end{Bmatrix}}_{\mathbf{h}} \tag{28}$$

with $0 \leq \gamma_L^{(E)} \perp \Lambda_L^{(E)} \geq 0$ and $0 \leq \gamma_R^{(E)} \perp \Lambda_R^{(E)} \geq 0$.

### 3.1 Mechanical model

In this part, the numerical simulation of one classical model for studying stick-slip vibration is used as an example for a 1-dimensional frictional contact problem. Figure 8 shows a one-degree of freedom slider-belt model with friction being the primary factor affecting the dynamics of the system.

Although the model looks simple, it can involve nonlinear or discontinuous factors that produce various behaviour. Thus, it is a general model for mechanism study and method testing. In this model, a point mass $m$ is held by a linear spring and viscous damper, supported by a rigid belt moving at a constant velocity $v_0$. $k_0$ is the spring stiffness, $c_0$ is the damping, and $F_n$ is the normal compression force. Coordinate $x$ describes the horizontal displacement. At the contact interface, the friction force is assumed to be proportional to the normal force.

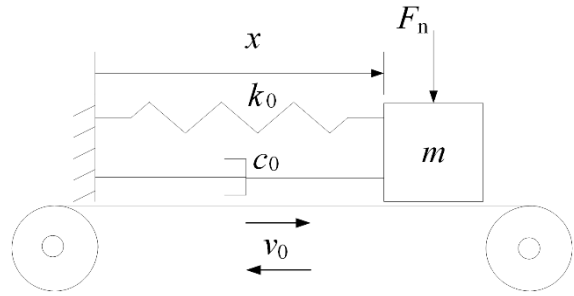

Figure 8. Model I: the single mass-belt system with friction.

The vertical position of this model is a constant value assumed to be 0. $\lambda_N = F_n$ is known in advance. The constraint function in the tangential direction are expressed as $g_T = \dot{x} - v_0 t$. According to the definitions, one gets: $W_T = 1$, $\bar{w}_N = \hat{w}_N = \bar{w}_T = 0$, $\hat{w}_T = -v_0$.

Figure 9 illustrates the two kinds of Coulomb-Stribeck laws of friction that are used for the following simulation in the forms below:

$$\mu = \frac{\mu_s}{1+\delta|v_r|}, \quad \mu = \mu_s + (\mu_s - \mu_d)e^{-\alpha|v_r|} \tag{29}$$

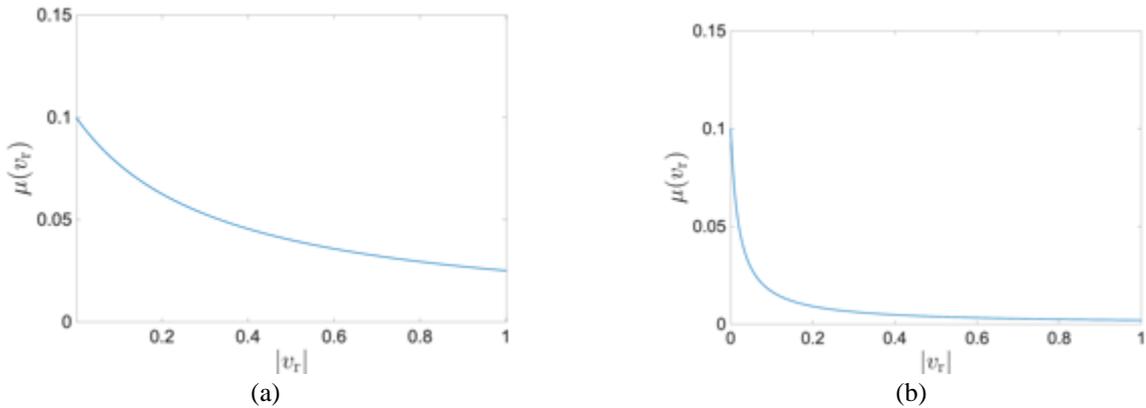

Figure 9. The friction law with different decreasing rates: (a) $\mu_s = 0.1$ and $\delta = 1$ (b) $\mu_s = 0.1$ and $\delta = 10$.

## 3.2 PINN scheme

Section 2 has introduced the basic idea of utilising PINN for solving LCP and nonsmooth dynamic problems. When the system is nonlinear/nonsmooth, the requirement for accuracy is stricter as the results of the next step are sensitive to the results of the previous step. To explore the applicability and accuracy of the new PINN strategy, two specific PINN frameworks for the single/dual PINN strategy are used for stick-slip simulation. Only the friction force is calculated from the conventional LCP algorithm involved in the single PINN framework or the LCP PINN part in the dual PINN framework. The LCP formulations are now Eq.(28).

## 3.3 Numerical Simulation

For the single-point frictional oscillator with stick-slip motion, the numerical methods, for example, the smoothing method or the bisection method, can provide very accurate results. Here, the switching model method to give ground truth results. But this method is impractical for multi-point frictional oscillators considering nonsmoothness factors such as stick-slip, loss of contact or impact. In this section, the mix-level time-stepping method presented in Section 2 is used to simulate stick-slip, which is called the conventional method. In the numerical simulations, system parameters are $m=1$, $k=1$, $c=0$.

### 3.3.1 *Friction law with a slowly decreasing rate*

Firstly, the PINN frameworks are examined in a situation when the friction force gently decays with the relative velocity. The nonlinear friction law that is employed has been given in Eq. (29) and depicted in Figure 9. Figure 10 shows the results of the phase plot of the vibration calculated by two PINN strategies. The black line with the triangle markers is about the ground truth results calculated by the switching method. The green line with the square markers is about the result of the single PINN framework, and the blue dashed line with a circle mark is about the result of the dual PINN framework. The straight line in the phase plot indicates stick motion, and the curve represents the slip motion. The vibration of the system alternates between the stick phase and the slip phase. The comparison in Figure 10 shows that the results from both the single PINN framework and dual PINN framework are identical to the ground truth results. This indicates that the idea of employing nonsmooth dynamics theory for the multi-point contact problem in a neutral network works very well in dealing with nonlinear stick-slip vibration. The $4^{th}$-order PINN and $10^{th}$-order PINN give very close results. Thus, only one group of results are provided.

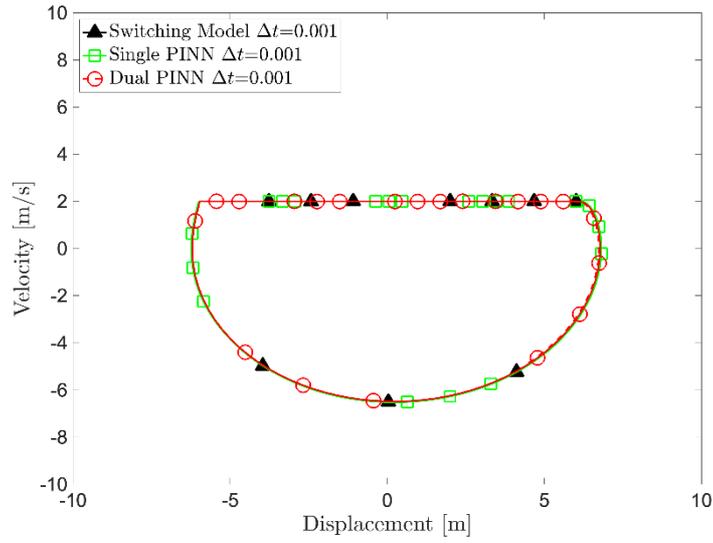

Figure 10. The phase plot of stick-slip vibration.

The dynamic behaviour of nonlinear systems with stick-slip or loss of contact (that would appear in 2-DoF models) depends on the initial condition. Hence, to capture the exact transition point which serves as the initial conditions for the next motion phase is one of the key points in accurate numerical calculation. Time-stepping method does not require the transition check or event check, while the price is a sufficiently small time step is required for high accuracy. Figures 11-12 illustrates the comparisons between conventional time-stepping methods and two PINN frameworks when the time step $\Delta t$=0.01.

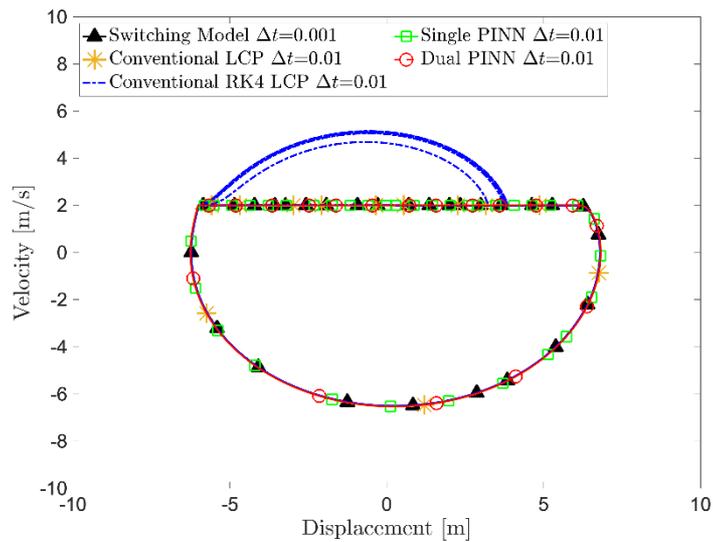

Figure 11. The phase plot of stick-slip vibration when Δ$t$=0.01.

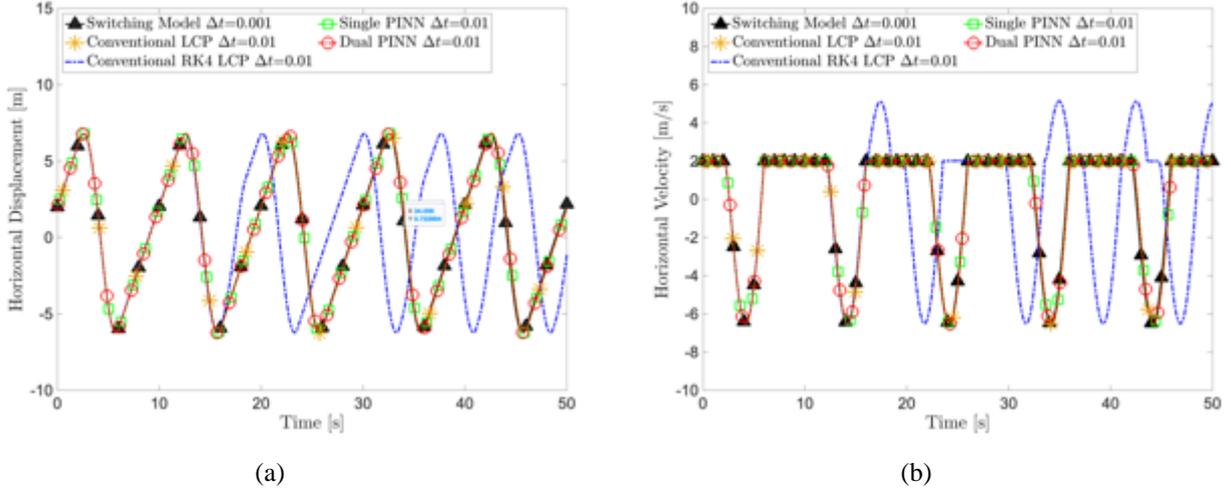

(a)                      (b)

Figure 12. Time history of the displacement and the velocity Δ$t$=0.01.

The calculation of the conventional LCP method and the conventional RK4 LCP method are based on a numerical integration method. The latter is a high-order numerical method. Figure 11 shows that when Δ$t$=0.01, the conventional LCP method could still predict the transitions between stick and slip motion. However, apparently higher-order RK4 LCP method, in this case, is not stable enough to predict the correct stick-slip. Notably, both the single and Dual PINN frameworks are high-order methods and could produce the correct stick-slip limit cycle. Furthermore, from Figure 12, it is known that the accuracy of PINN frameworks in both the displacement and the velocity level is quite good. The conventional RK4 LCP method could initially predict the correct results until the failure of detecting the second transition from slip to stick around 18s (illustrated in Figure 12 (b)). As high-order algorithms, the PINN frameworks are more reliable.

In some dynamic problems, the difference between the time histories of the vibration calculated by different methods is not remarkable. The frequency spectrum shown in Figure 13 indicates that the system with stick-slip motion has one primary vibration frequency, one doubling frequency and one tripling frequency. From the frequency point of view, when a method (Conventional RK4 LCP method) is incapable of simulating correct stick-slip transitions, it not only predicts distorted time history responses but also provides incorrect vibration frequencies. Stick-slip vibration has a higher requirement for numerical methods.

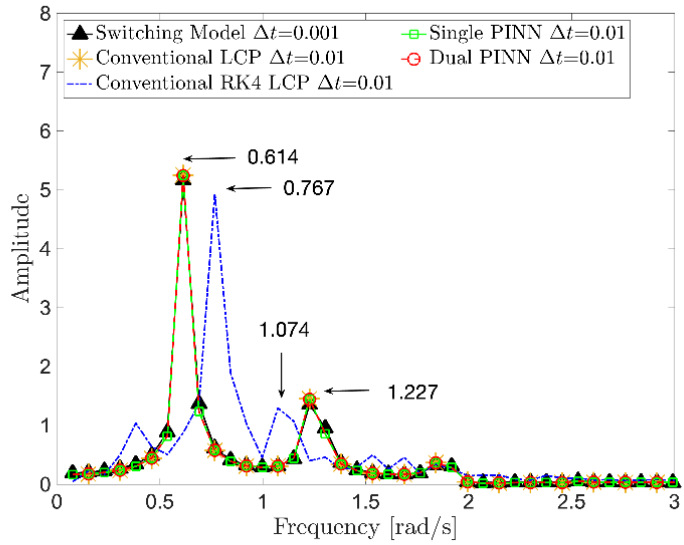

Figure 13. The frequency of stick-slip vibration Δ*t*=0.01.

When the time step increases to Δ*t*=0.08, Figure 14 shows that the two PINN frameworks could maintain the correct prediction of stick-slip vibration. However, the two conventional methods could not give accurate results. Figure 14 (b) is the time history of the velocity, in which the constant velocity means that the system is in the stick phase. Otherwise the slider is slipping. Figure 14 (b) explains the reason for the failure of the conventional methods. With the large time step, the conventional method cannot identify the transition instants between stick and slip motion.

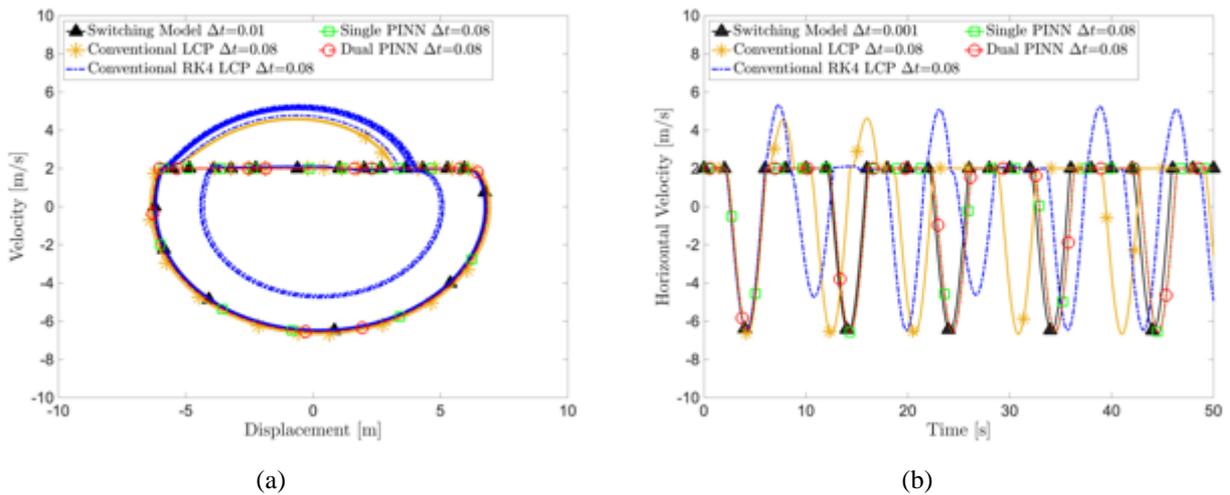

(a)        (b)

Figure 14. The time response of stick-slip vibration when Δ*t*=0.08.

### 3.3.2  *Friction law with a sharply decreasing rate*

In this section, the simulations using the friction law shown in Figure 9 (b) are conducted. The sharp decrease of the friction force with the velocity near zero velocity reflects the friction behaviour of many real frictional pairs. This kind of friction law causes more difficulties to numerical calculations. On the one hand, using a smaller time step could ensure the detection of stick-slip transitions which has been known as a key factor for the accuracy of nonsmooth vibration. On the other hand, the time step could not be too small, considering the expenses of numerical calculation. In the following examples, the neural network that is used in the PINN for dynamic simulation are a 7-layers feedforward neutral network in which each layer contains 20 neutrons. The network used in the PINN for LCP are a 3-layers feedforward neutral network in which each layer contains 5 neutrons.

Figure 15 show the results of different methods at the time step that could predict correct stick-slip vibration. As illustrated in Figure 15 (b), the conventional high-order method needs a very small time step comparing with the PINN frameworks. The dual PINN framework needs a time step that is 50 times larger than the Conventional RK4 LCP method. Although the advantage of the Single PINN framework in terms of the time step is not as outstanding as the dual PINN framework, it could predict stick-slip vibration accurately when the time step is not too small.

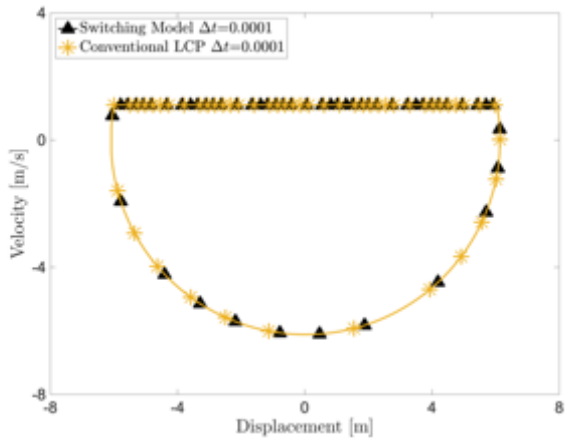
(a)

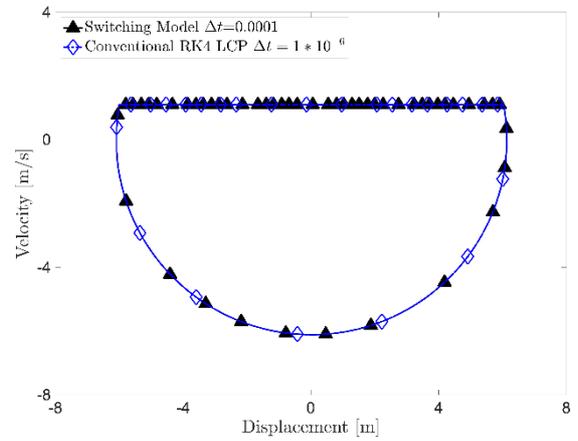
(b)

23### 3.3.2  *Friction law with a sharply decreasing rate*

In this section, the simulations using the friction law shown in Figure 9 (b) are conducted. The sharp decrease of the friction force with the velocity near zero velocity reflects the friction behaviour of many real frictional pairs. This kind of friction law causes more difficulties to numerical calculations. On the one hand, using a smaller time step could ensure the detection of stick-slip transitions which has been known as a key factor for the accuracy of nonsmooth vibration. On the other hand, the time step could not be too small, considering the expenses of numerical calculation. In the following examples, the neural network that is used in the PINN for dynamic simulation are a 7-layers feedforward neutral network in which each layer contains 20 neutrons. The network used in the PINN for LCP are a 3-layers feedforward neutral network in which each layer contains 5 neutrons.

Figure 15 show the results of different methods at the time step that could predict correct stick-slip vibration. As illustrated in Figure 15 (b), the conventional high-order method needs a very small time step comparing with the PINN frameworks. The dual PINN framework needs a time step that is 50 times larger than the Conventional RK4 LCP method. Although the advantage of the Single PINN framework in terms of the time step is not as outstanding as the dual PINN framework, it could predict stick-slip vibration accurately when the time step is not too small.

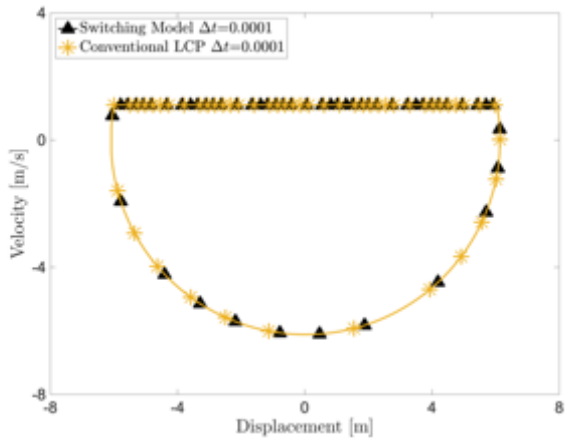
(a)

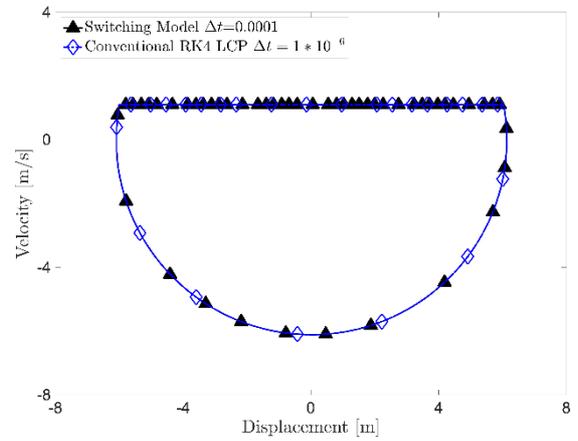
(b)

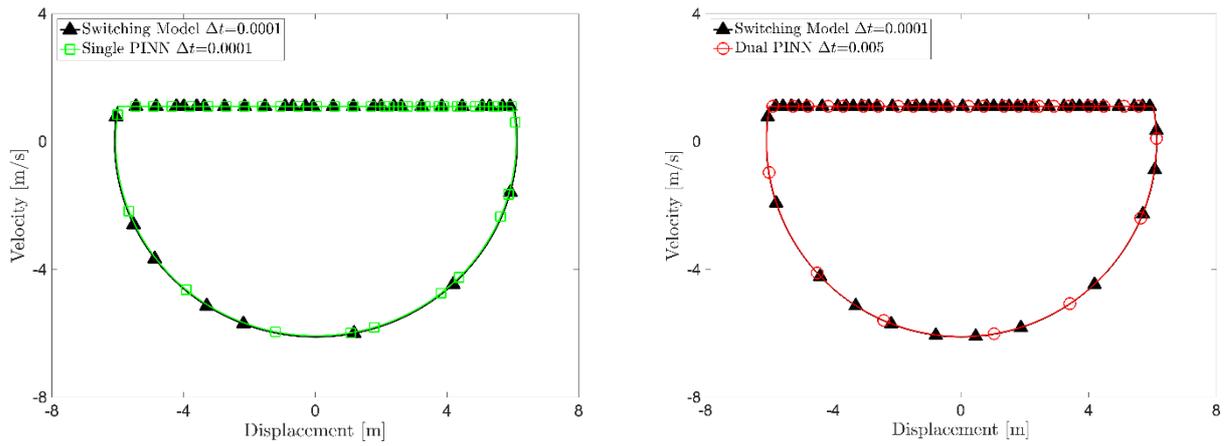

Figure 15. The correct phase plot of stick-slip by different methods: (a) conventional LCP $\Delta t$=0.0001; (b) Conventional RK4 LCP when $\Delta t$=10$^{-6}$; (c) Single PINN when $\Delta t$=0.0001; (d) Dual PINN when $\Delta t$=0.005.

Figure 16 compares the time history results of the different methods based on LCP strategy for multi-contact systems when $\Delta t$=0.005. Figure 16 (b) shows that the Single PINN framework and the two conventional methods at the same time step fail to switch to stick motion after the slider starts to slide on the belt. Consequently, the wrong displacement track begins, as shown in Figure 16 (a). The above results indicate the importance of the numerical method in correct state detection and the excellent performance of Dual PINN.

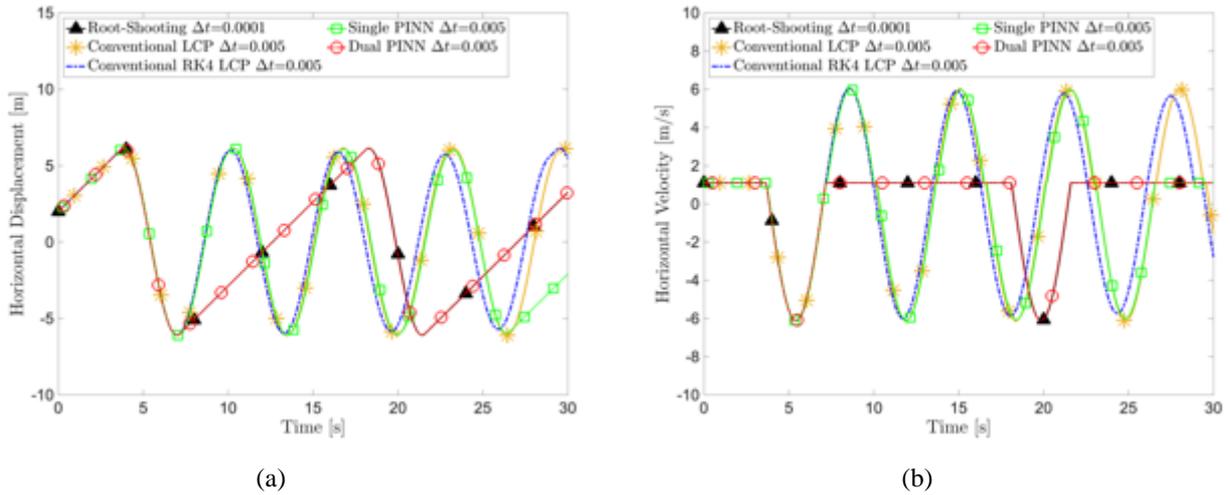

(a) (b)

Figure 16. The time response of stick-slip vibration with friction law when $\Delta t$=0.005: (a) the time history of the displacement; (b) the time history of the velocity.

# 4 Two-dimensional nonsmooth FIV with spring contact

## 4.1 Mechanical model

Figure 17 is one classic and minimal model that may show stability bifurcation, known as friction-induced mode-coupling instability. It is a 2-DoF model consisting of a mass-spring-damper part as a slider and a rigid moving belt. In the horizontal direction, the mass $m$ is held by a spring $k_1$ and a damper $c_1$. In the vertical direction, $m$ is linked with a ground damper $c_2$. Additionally, $m$ is constrained by an inclined spring $k_3$, which would cause an asymmetry stiffness that accounts for mode-coupling instability. Between the mass and the belt, the contact stiffness is $k_2$. An assumed massless slider is in contact with the moving belt. A preload $F_p$ is applied to bring the slider into contact with the belt before the belt starts to move. The friction at the interface is assumed to follow Coulomb-Stribeck friction law. The sliding friction force $F_T$ is proportional to the normal contact force.

As it has been known, the coefficient of friction $\mu$ affects the natural frequencies of the 2-DoF system, which has been investigated in Refs. [35, 36]. With the increase of $\mu$, two natural frequencies of the system get closer to each other and eventually merge, resulting in instability. As horizontal vibration grows to a certain level, the vertical motion occurs and the contact between the slider and the belt is lost as there is not a connection between the two parts; and subsequently they may get into contact again. The repeated transition of the states in the normal direction is called the separation-reattachment phenomenon. Additionally, when Coulomb-Stribeck friction law is considered, the slider would undergo horizontal stick-slip vibration at the interface. The unstable vibration of the system could be aroused by stick-slip instability. Both types of nonsmooth phenomena can be present this model, which is unlike the one-degree-of-freedom model in Section 3.

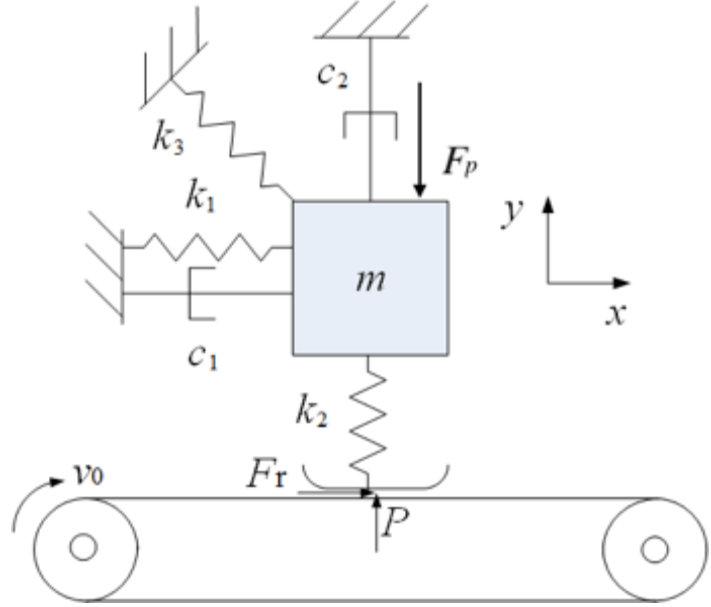

Figure 17. 2-DoF slider-belt model with dry friction

in certain conditions, the vibration regime alternates in stick, slip and separation phases. The FIV model would be a nonlinear and nonsmooth system, and a time-domain analysis is necessary, which is different from previous research [35]. This part uses PINN frameworks to simulate the transient dynamic response, considering the nonsmoothness in the normal direction and tangential direction.

The general equation of motion for unilateral contact with friction is given in Eq.(6). The unknow vector of this example is $\mathbf{q}=[x \ y]^T$. For this model, $\mathbf{M}=\begin{bmatrix} m & 0 \\ 0 & m \end{bmatrix}$, $\mathbf{h}=\begin{Bmatrix} -c_1\dot{x}-(k_1+k_3/2)x-k_3y/2 \\ -c_2\dot{y}-k_3x/2+k_3y/2+F_p \end{Bmatrix}$.

The parameter values that are used in the following simulations are $m=5$, $k_1=1000$, $k_3=600$, $k_c=500$, $c_1=c_2=0$.

The normal constraint function for the 2-DoF model is $g_N=y$. The tangential constraint function is $\gamma_T=\dot{x}-v_0$. Thus, according to definitions in Section 2.2, one can get $\mathbf{W}_N$, $\mathbf{W}_T$, $\hat{\omega}_N$ and $\hat{\omega}_T$ are:

$$\mathbf{W}_N=\left(\frac{\partial g_N}{\partial \mathbf{q}}\right)^T=[0 \ 1]^T, \ \mathbf{W}_T=\left(\frac{\partial g_T}{\partial \mathbf{q}}\right)^T=[1 \ 0]^T, \ \hat{\omega}_N=0 \ \text{and} \ \hat{\omega}_T=-v_0.$$

### 4.2 Advanced PINN scheme for the two-dimensional nonsmooth problem

The single PINN strategy and dual PINN strategy have been given in Section 2.3.3 and Section 2.3.4. As the vibration with 2D nonsmoothness is more complex, to improve the accuracy of the single/ dual PINN framework, an interpolation technique is used. According to the order of the numerical integration method, the force vector $\boldsymbol{\lambda}_T^{(i)}$ and $\boldsymbol{\lambda}_N^{(i)}$ in conventional methods are now transformed to $\boldsymbol{\lambda}_{T\text{-RK}}^{(i)}$ and $\boldsymbol{\lambda}_{N\text{-RK}}^{(i)}$ matrices based on the following equations:

$$\boldsymbol{\lambda}_{T\text{-RK}}^{(i)} = \boldsymbol{\lambda}_T^{(i-1)} + \mathbf{s}\bullet\left(\boldsymbol{\lambda}_T^{(i)} - \boldsymbol{\lambda}_T^{(i-1)}\right) \tag{30}$$

$$\boldsymbol{\lambda}_{N\text{-RK}}^{(i)} = \boldsymbol{\lambda}_N^{(i)} + \mathbf{s}\bullet\left(\boldsymbol{\lambda}_N^{(i)} - \boldsymbol{\lambda}_N^{(i-1)}\right) \tag{31}$$

in which $\mathbf{s}$ is the vector formed by the time segment calculated based on the coefficient $c_k$ ($k=1,2,\ldots,q$) of the $q^{\text{th}}$-order RK method. Then $\boldsymbol{\lambda}_{T\text{-RK}}^{(i)}$ and $\boldsymbol{\lambda}_{N\text{-RK}}^{(i)}$ are passed into the physics part of the PINN for dynamic response $\boldsymbol{\lambda}_T^{(i)}$ and $\boldsymbol{\lambda}_N^{(i)}$ in Eq. (c) in Figures 5 and 6. The new methods are referred to as the advanced Single PINN and the advanced Dual PINN in the following. The solution procedure of the two advanced PINN frameworks is shown in Figure 18.

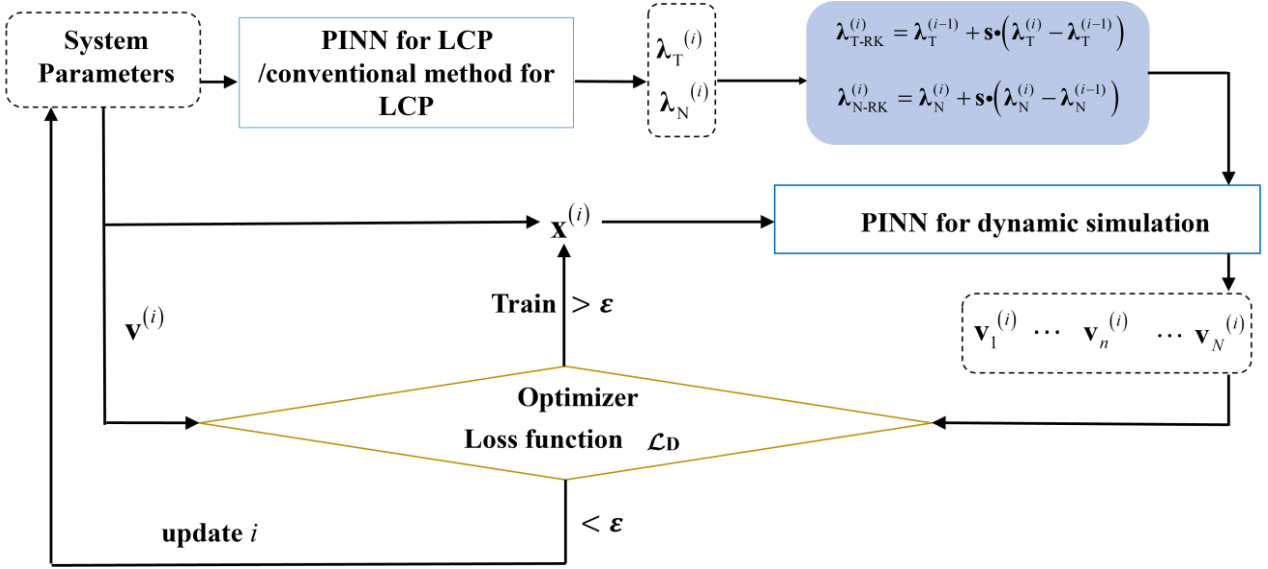

Figure 18. Framework of the advanced Single/Dual PINN for the 2D nonsmooth problem.

## 4.3 Numerical simulation

Although stick-slip and separation-reattachment are allowed in this problem, the occurrence of these phenomena depends on the system parameters. As mode-coupling instability may happen, the critical coefficient of friction $\mu_c$ at bifurcation can be obtained through the complex eigenvalue analysis. Figure 19 shows the change of the real and imaginary parts of the eigenvalues (of the system during sliding contact) with the coefficient of friction for Example 1 (see section 4.3.1). One can observe that instability occurs at $\mu_c=0.83$. In the following, two examples with parameters $\mu=0.4$ and $\mu=1.2$, respectively, are investigated.

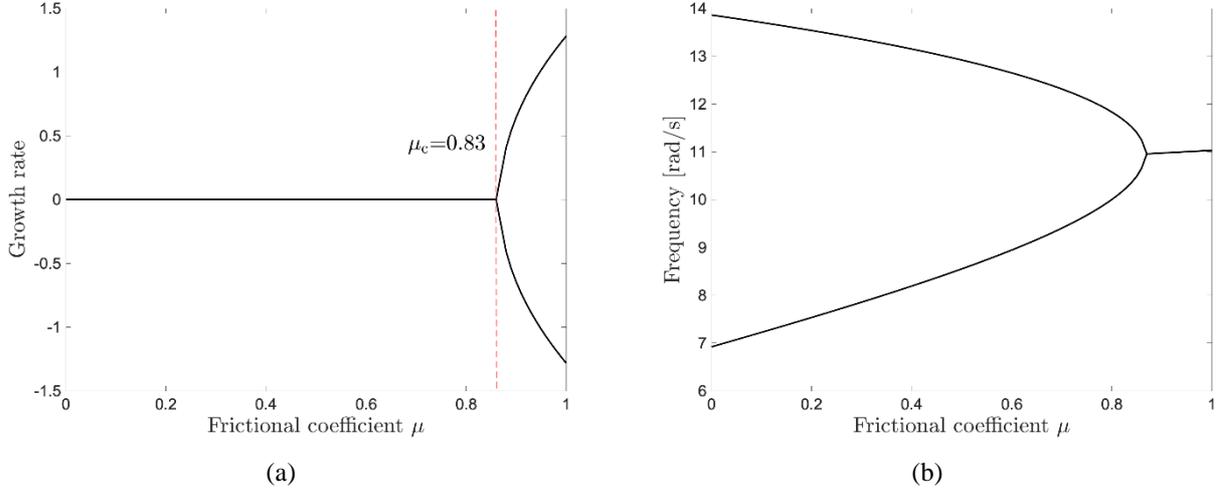

(a)  (b)

Figure 19. The evolution of the real part (left) and imaginary part (right) of the eigenvalue with the friction coefficient.

### 4.3.1    *Example 1:*

Based on the advanced single PINN and dual PINN schemes, numerical simulations are carried out. The initial condition is $x(t=0)=y(t=0)=-10$, and $v_x(t=0)=v_y(t=0)=v_0$. The size of the neutral network in the following examples is identical to the problem in Section 3. Figure 20 shows the results of the advanced PINN frameworks with the root-shooting strategy. The results of the root-shooting method are used as the ground truth results. In Figures 20 and 21, the solid and hollow squares are for the advanced single PINN method, when a $4^{th}$-order and a $10^{th}$-order integration formulations are used in the PINN for dynamic simulation. The solid triangles and hollow circles are for the advanced dual PINN method when a $4^{th}$-order and a $10^{th}$-order integration formulations are involved.

Figure 20 shows the time history of the contact force. In this example, the contact force alternates between positive values, which represent contact between the slider and the belt, and zero value, which represents the loss of contact. Thus, separation and reattachment happen during the vibration, which brings the vertical nonsmoothness. From the results, one can see that both the advanced single PINN and the advanced dual PINN framework can produce very good results for a problem when repeated separation-reattachment events happen. As illustrated in the enlarged plot of the contact force (Figure 20 (b)), when the $4^{th}$-order PINN is used, the result calculated by the dual PINN framework is not as good as that of the single PINN. When the $10^{th}$-order integration method is used, the accuracy of the dual PINN framework (red circles) can be improved from $4^{th}$-order dual PINN (blue triangle). Similarly, the accuracy of the single PINN framework can be also improved by using a high-order integration method.

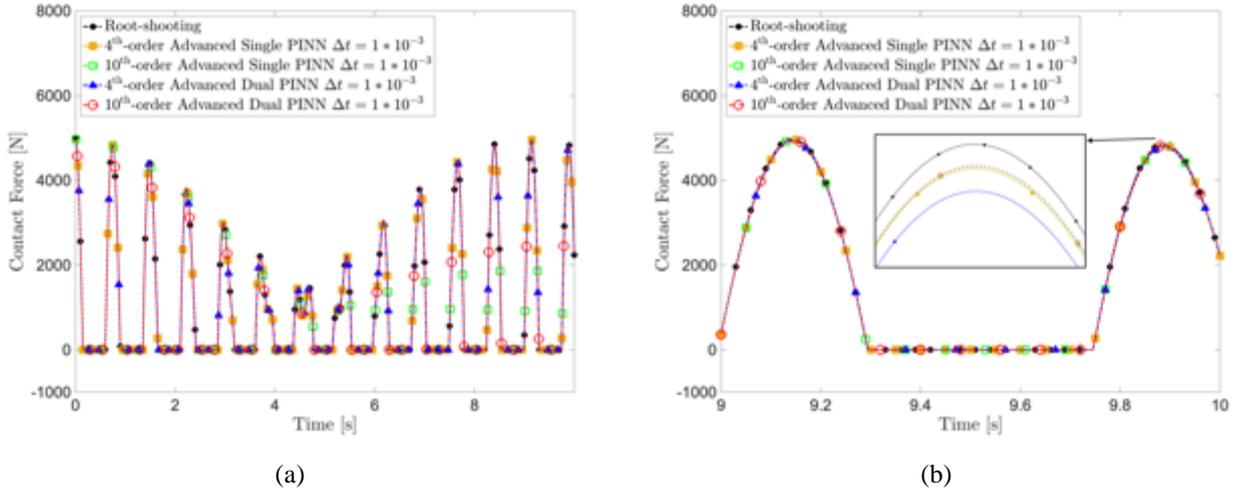

Figure 20. The time history of the contact force. (a) $t$=[0~10]s; (b) $t$=[9~10]s.

Figure 21 illustrates that the displacement and velocity responses calculated by the two advanced PINN strategies are very accurate compared with the root-shooting results. As shown by the time history of the velocity given in Figure 21 (c)-(d), stick-slip vibration does not appear in this case. These results enable comparisons between the PINN strategies that employ the dynamic equations with unilateral contact and the ground truth results, which are calculated by the rooting-searching method for the single-point contact problem. In the following, the comparisons between the conventional numerical methods for the multi-point contact problem with the PINN strategies are made.

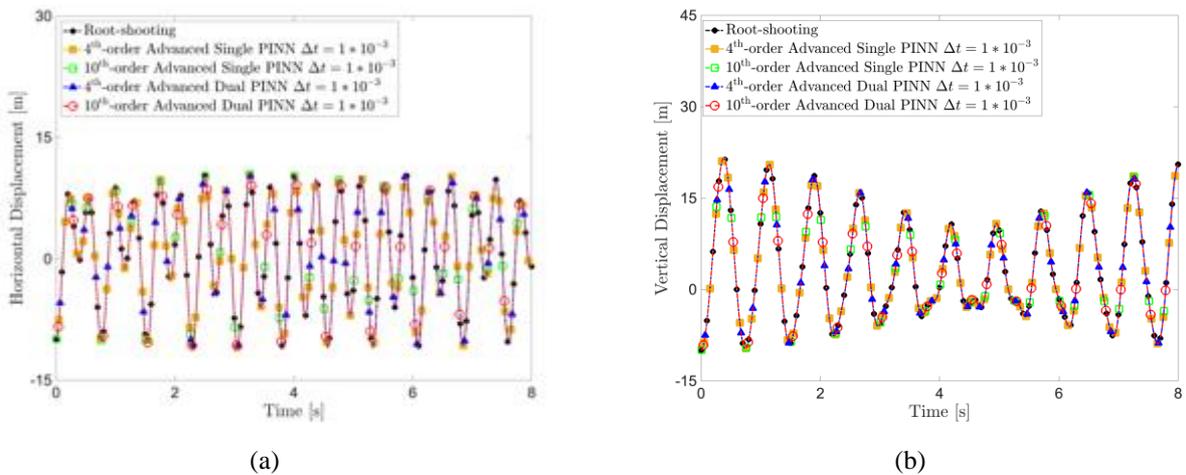

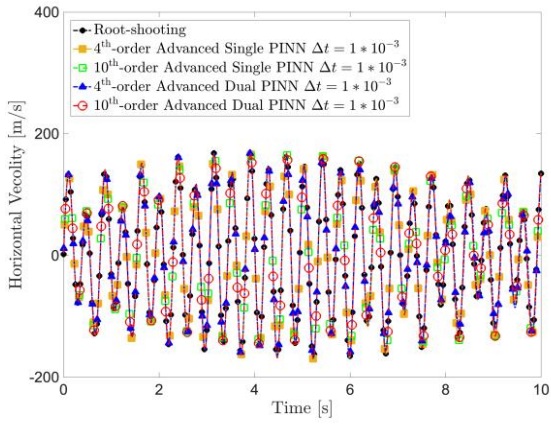
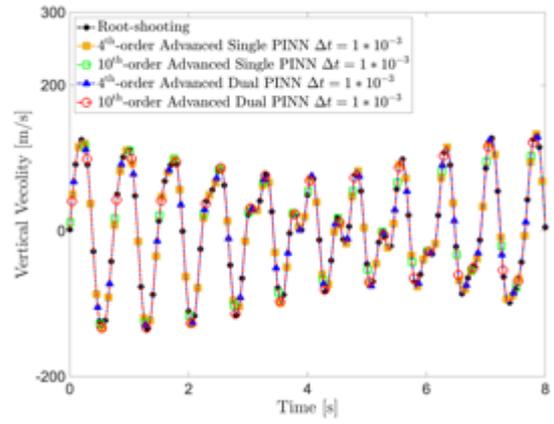

(c)                  (d)

Figure 21. The time history of displacement and velocity: (a) horizontal displacement vs time; (b) vertical displacement vs time; (c) horizontal velocity vs time; (d) vertical velocity vs time.

Table 2 shows the root mean squares (RMS) of the contact force, the displacement and the velocity calculated by different methods. The difference between the conventional LCP method and the RK4 LCP method is that the former uses Morue's algorithm, and the latter adopts the idea of $4^{th}$-order RK. Single PINN and Dual PINN are the methods without the interpolation that are introduced in Sections 2.2.3 and 2.2.4. To give a clearer illustration of the data in Table 2, Figure 22 shows the RMS error of the horizontal and vertical displacement. By using the results of the root-shooting method as the reference value, the absolute errors of the corresponding RMS results of the six methods for nonsmooth simulation of multi-point contact problem are calculated. One can find that: (1) with the interpolation technology, the advanced PINN frameworks improve the accuracy of the PINN framework. (2) compared with the conventional methods, the advanced PINN frameworks have the advantage in accuracy when the same time step is used ($\Delta t$ =0.001). The accuracy of the conventional methods could be improved when the time step is smaller ($\Delta t$ =0.0005), which means lower efficiency. As shown by the velocity data listed in Table 2, the proposed advanced single and dual PINN methods can produce accurate results.

Table 2. RMS results of different numerical methods.

| Method \ RMS | Contact force | Horizontal displacement | Vertical displacement | Horizontal velocity | Vertical velocity |
|---|---|---|---|---|---|
| Root-Shooting Method | 1790 | 6.298 | 9.647 | 94.883 | 75.042 |

| | | | | | | |
|---|---|---|---|---|---|---|
| Conventional LCP method | $\Delta t=10^{-3}$ | 1716 (4.13%) | 6.195 (1.64%) | 9.322 (3.37%) | 93.640 (1.31%) | 72.544 (3.33%) |
| | $\Delta t=5\times10^{-4}$ | 1780 (0.56%) | 6.285 (0.21%) | 9.604 (0.45%) | 94.721 (0.17%) | 74.709 (0.44%) |
| Conventional RK4 method | $\Delta t=10^{-3}$ | 1752 (2.12%) | 6.245 (0.84%) | 9.480 (1.73%) | 94.253 (0.66%) | 73.755 (1.72%) |
| | $\Delta t=5\times10^{-4}$ | 1785 (0.28%) | 6.292 (0.09%) | 9.625 (0.23%) | 94.801 (0.09%) | 74.878 (0.22%) |
| 10th-order Single PINN | $\Delta t=10^{-3}$ | 1751 (2.03%) | 6.245 (0.84%) | 9.478 (1.67%) | 94.254 (0.66%) | 73.749 (1.72%) |
| 10th-order Dual PINN | $\Delta t=10^{-3}$ | 1752 (2.12%) | 6.249 (0.78%) | 9.485 (1.74%) | 94.320 (0.59%) | 73.794 (1.66%) |
| 10th-order advanced single PINN | $\Delta t=10^{-3}$ | 1785 (0.28%) | 6.291 (0.11%) | 9.626 (0.21%) | 94.977 (0.10%) | 74.968 (0.10%) |
| 10th-order advanced dual PINN | $\Delta t=10^{-3}$ | 1786 (0.22%) | 6.296 (0.03%) | 9.632 (0.15%) | 94.984 (0.11%) | 74.983 (0.08%) |

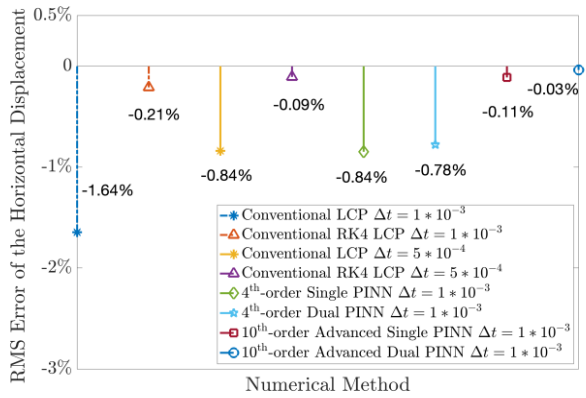
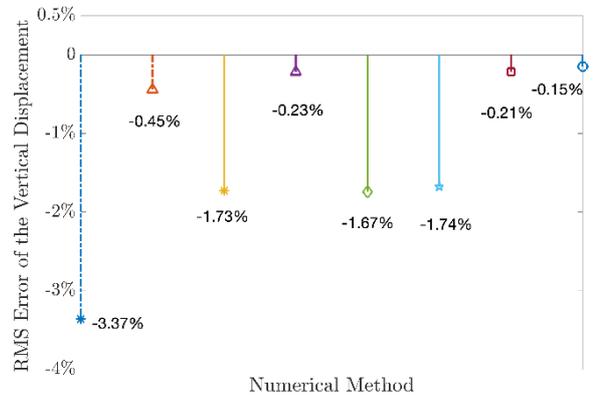

(a)　　　　　　　　　　　　　　　　　(b)

Figure 22. The RMS error of the displacement response: (a) horizontal displacement; (b) vertical displacement.

### 4.3.2 *Example 2*

Figures 23 and 24 show the response of the contact force and the horizontal velocity when $\mu$=1.2 and the initial conditions are $x(t=0) = y(t=0) = -1$, and $v_x(t=0) = v_y(t=0) = v_0$. The first observation is that separation and reattachment happen in this example, as the contact force repeatedly alternants between positive values and zero representing in contact and loss of contact, respectively. Then from Figure 24, it can be seen that stick-slip vibration in the horizontal direction occurs as well. More clearly, Figure 24 (b) shows that when both normal and tangential nonsmoothness are involved in a problem, the state change is unpredictable. Figure 23 shows whether the advanced single PINN or advanced dual PINN framework can accurately detect the separation-reattachment events. The zoomed-in view is presented in Figure 23 (b). Compared with the results of the $10^{th}$-order advanced single and dual PINN cases, adding one PINN could lower the accuracy, which also happens in Example 1. However, when a higher order numerical integration scheme is used, the accuracy of both single and dual PINN frameworks is improved, and it is not a big concern. As shown in Figure 24, stick-slip vibration occurs from around 4.6s. Consistently with the contact force results, when the order of the numerical method used is ten, accurate stick-slip vibration can be produced by both the advanced Single and dual PINN frameworks. Even though the dynamic response is highly nonlinear, PINN frameworks work quite well.

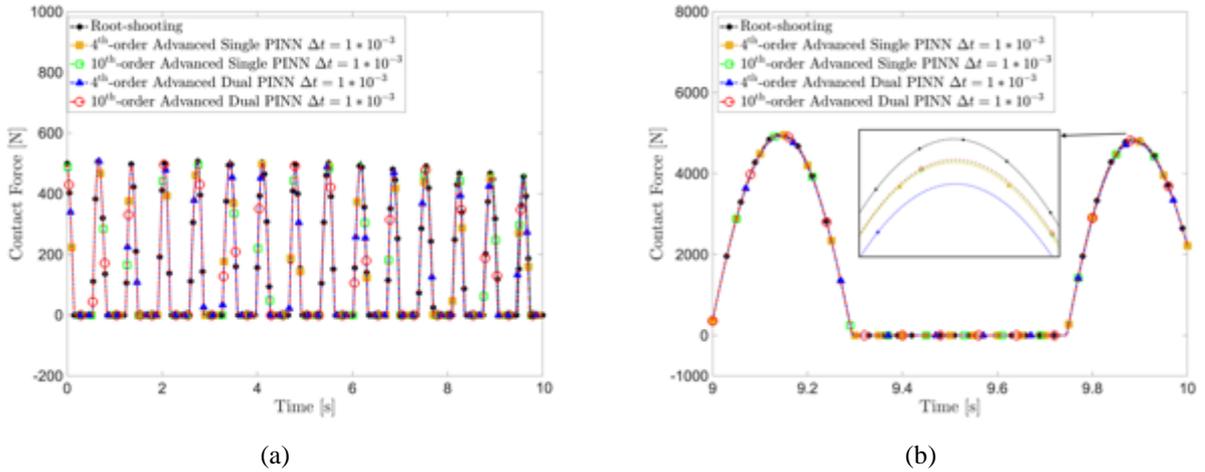

Figure 23. The time history of the contact force: (a) $t$=[0~10] s; (b) $t$=[8.7~9.7]s.

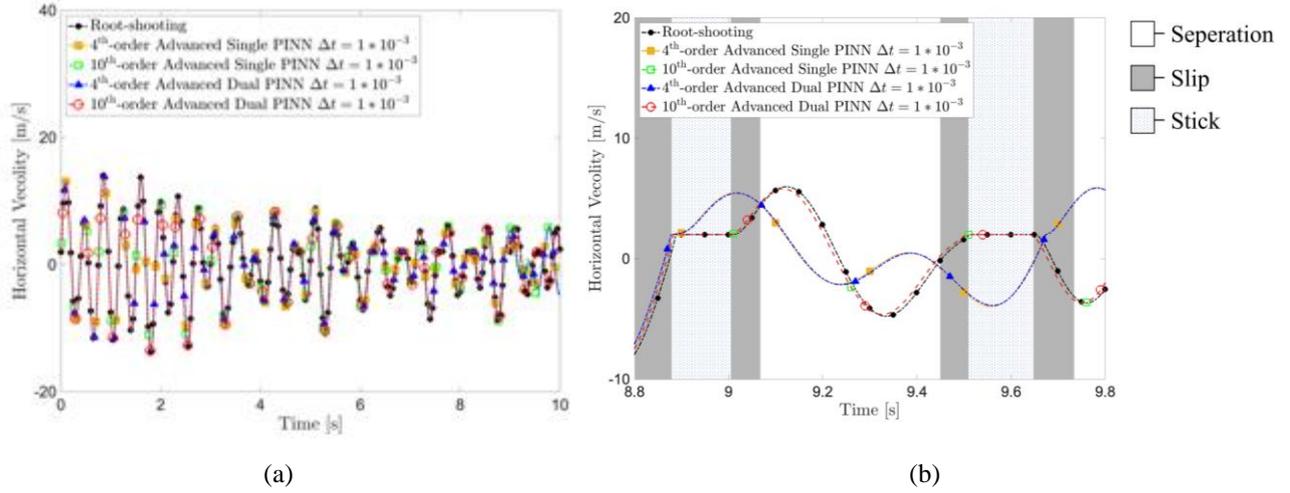

(a)                                                (b)

Figure 24. The time history of the velocity. (a) $t$=[0~10] s; (b) $t$=[8.7~9.7]s, in which the dotted zone represents the sticking state, the grey zone represents the slipping state, the white zone represents the separation state.

In the following, the results of the new PINN frameworks and the conventional methods that all nonsmooth calculations of multi-point contact systems are compared. In Table 3, numerical calculations of the two conventional methods are carried out with different time steps ($\Delta t$ =0.001, $\Delta t$ =0.0005, and $\Delta t$ =0.0001). × denotes the failure of detection of stick-slip transitions, which in the view of nonsmooth simulations is unacceptable. When $\Delta t$ =0.0001, conventional methods could give satisfied results. With respect to the PINN frameworks, when the 4$^{th}$-order methods are used, a smaller time step is needed as when $\Delta t$ =0.001 would not correctly predict stick-slip transitions at 8.88 second (see Figure 24(b)). Furthermore, the 10$^{th}$-order single and dual PINN frameworks can make an accurate transition detection at $\Delta t$ =0.001, which is 10 times larger than the time step of the conventional methods, and the accuracy of the overall time- domain response is high.

Table 3. RMS results of different numerical methods. Cross (×) denotes results that fail to present correct stick-slip transitions

| Method | RMS | Contact force | Horizontal displacement | Vertical displacement |
|---|---|---|---|---|
| Root-Shooting Method | | 224.449 | 0.3997 | 0.7629 |
| Conventional LCP method | $\Delta t$=10$^{-3}$ | × | × | × |
| | $\Delta t$=5×10$^{-4}$ | × | × | × |
| | $\Delta t$=10$^{-4}$ | 222.331 (0.94%) | 0.3980 (0.43%) | 0.7530 (1.30%) |
| | $\Delta t=10^{-3}$ | × | × | × |
| Conventional RK4 method | $\Delta t=5\times10^{-4}$ | 224.396 (0.02%) | 0.7603 (90.22%) | 0.7603 (0.34%) |
| | $\Delta t=10^{-4}$ | 224.221 (0.10%) | 0.3996 (0.03%) | 0.7619 (0.13%) |
| 4$^{th}$-order Advanced Single PINN | $\Delta t=10^{-3}$ | × | × | × |
| | $\Delta t=5\times10^{-4}$ | 222.424 (0.90%) | 0.3977 (0.50%) | 0.7537 (1.21%) |
| 4$^{th}$-order Advanced Dual PINN | $\Delta t=10^{-3}$ | × | × | × |
| | $\Delta t=5\times10^{-4}$ | 213.058 (5.08%) | 0.3976 (0.53%) | 0.7538 (1.19%) |
| 10$^{th}$-order Advanced Single PINN | $\Delta t=10^{-3}$ | 221.828 (1.17%) | 0.3970 (0.68%) | 0.7512 (1.53%) |
| 10$^{th}$-order Advanced Dual PINN | $\Delta t=10^{-3}$ | 221.835 (1.16%) | 0.3969 (0.70%) | 0.7511 (1.55%) |

Through the above numerical analysis of examples 1 and 2, one can conclude that when determining the vibration of a system considering separation-reattachment and stick-slip, the numerical methods and the time step must be chosen very carefully. The error in predicting nonsmooth vibration comes not only from the discernible error accumulation due to the numerical integration, but also from the unpredictable failure of detecting nonsmooth transitions, which is a fatal issue. The new PINN frameworks proposed in this work are shown to be well capable of dealing with nonsmoothness in two dimensions (the normal and tangential directions). Moreover, compared with conventional methods, the advanced single and deal PINN frameworks are less dependent on the time step length to give good predictions of the stick-slip and separation-reattachment transitions. Because PINN frameworks rely on the functions form conventional physical equations, in which the numerical error from the traditional numerical methods can be offset through the training of the neural network. The accuracy of PINN frameworks can be improved in two ways: decreasing the time-step and increasing the order of the numerical integration formulations in PINN frameworks.

## 5 Conclusions

This work proposes a novel idea for simulating the vibration of frictional systems with multiple contact points by employing the physics-informed neural networks (PINN) established by the authors. Based on that, general PINN frameworks for nonsmooth dynamic analysis for complex systems with friction are proposed. Firstly, a tailored PINN based on the mathematical expressions of linear complementary problem (LCP) is proposed. Then, the PINN for the dynamic simulation of complex systems based on numerical integration formulations of the dynamic equations is established. Furthermore, based on the two new PINN frameworks, four PINN methods to replace conventional numerical methods for solving the vibration of frictional systems with multiple contact points are proposed. They are the single PINN framework, the dual PINN framework, and the advanced single/dual PINN framework, respectively.

The applications of these frameworks in solving the direct contact problem with only the stick-slip vibration (1-D problem) and spring contact problem with both separation-reattachment and stick-slip (2-D problem) lead to the following findings: (1) all four PINN frameworks can accurately detect stick-slip and separation-reattachment transition events in both 1-D and 2-D nonsmooth problems; (2) Compared with the conventional numerical methods, the PINN frameworks do not necessitate a tiny time step. High-order PINN frameworks allow a larger time step with better accuracy and greater stability than the conventional methods; (3) the advanced single/dual PINN frameworks outperform the single/dual PINN frameworks in identifying state transitions, which is important for achieving high accuracy; (4) the accuracy of PINN strategies can be improved by simply increasing the numerical integration order. Overall, the developed new method is found be capable of accurately predicting the nonsmooth dynamic behaviour.

## Acknowledgements

The authors are grateful for the financial support from China Postdoc Science Foundation (No. 2019M652564) and the Songshan Laboratory Project (No: 221100211000-01). Support from the Australian Research Council to SM and YG is also gratefully acknowledged (FT180100338; IC190100020).